\documentclass[english]{article}
\usepackage[utf8]{inputenc}
\usepackage[T1]{fontenc}
\usepackage{listings}
\usepackage{babel}
\usepackage{amsmath}
\usepackage{graphicx}
\usepackage{subfigure}
\usepackage{fancyhdr}
\usepackage{color}
\usepackage{caption}
\usepackage{tabularx}
\usepackage{array,tabularx,booktabs}
\usepackage{multirow}
\usepackage[table, dvipsnames]{xcolor}
\usepackage{caption}
\usepackage{hyperref}
\usepackage{authblk}
\usepackage{booktabs}

\colorlet{punct}{red!60!black}
\definecolor{background}{HTML}{EEEEEE}
\definecolor{delim}{RGB}{20,105,176}
\colorlet{numb}{black}

\lstdefinelanguage{json}{
    basicstyle=\normalfont\ttfamily,
    numbers=left,
    numberstyle=\scriptsize,
    stepnumber=1,
    numbersep=8pt,
    showstringspaces=false,
    breaklines=true,
    frame=lines,
    backgroundcolor=\color{background},
    literate=
     *{0}{{{\color{numb}0}}}{1}
      {1}{{{\color{numb}1}}}{1}
      {2}{{{\color{numb}2}}}{1}
      {3}{{{\color{numb}3}}}{1}
      {4}{{{\color{numb}4}}}{1}
      {5}{{{\color{numb}5}}}{1}
      {6}{{{\color{numb}6}}}{1}
      {7}{{{\color{numb}7}}}{1}
      {8}{{{\color{numb}8}}}{1}
      {9}{{{\color{numb}9}}}{1}
      {:}{{{\color{punct}{:}}}}{1}
      {,}{{{\color{punct}{,}}}}{1}
      {\{}{{{\color{delim}{\{}}}}{1}
      {\}}{{{\color{delim}{\}}}}}{1}
      {[}{{{\color{delim}{[}}}}{1}
      {]}{{{\color{delim}{]}}}}{1},
}

\newlength{\subcolumnwidth}

\newcommand{\nextsubcolumn}[1][]{%
  \cr\noalign{\hfill}
  \if\relax\detokenize{#1}\relax\else\hsize=#1\setlength{\subcolumnwidth}{\hsize}\fi
}

\begin{document}

\title{A dataset to assess mobility changes in Chile following local quarantines}

\author{Luca Pappalardo\textsuperscript{1}, Giuliano Cornacchia\textsuperscript{2}, Victor Navarro\textsuperscript{3,5}, \\Loreto Bravo\textsuperscript{3,4}, Leo Ferres\textsuperscript{3,4}}
\maketitle

1. ISTI-CNR, Pisa, Italy; 2. Department of Computer Science, University of Pisa; 3. Faculty of Engineering, Universidad del Desarrollo, Santiago, Chile; 4. Telef\'onica R\&D Santiago, Chile; 5. Department of Astronomy, University of Chile, Chile


\begin{abstract}
Fighting the COVID-19 pandemic, most countries have implemented non-pharmaceutical interventions like wearing masks, physical distancing, lockdown, and travel restrictions. Because of their economic and logistical effects, tracking mobility changes during quarantines is crucial in assessing their efficacy and predicting the virus spread. Chile, one of the worst-hit countries in the world, unlike many other countries, implemented quarantines at a more localized level, shutting down small administrative zones, rather than the whole country or large regions. Given the non-obvious effects of these localized quarantines, tracking mobility becomes even more critical in Chile. To assess the impact on human mobility of the localized quarantines in Chile, we analyze a mobile phone dataset made available by Telef\'onica Chile, which comprises 31 billion eXtended Detail Records and 5.4 million users covering the period February 26th to September 20th, 2020. From these records, we derive three epidemiologically relevant metrics describing the mobility within and between comunas. The datasets made available can be used to fight the COVID-19 epidemics, particularly for localized quarantines' less understood effect.
\end{abstract}

\section{Background \& Summary}
\label{sec:introduction}
As of November 2020, the COVID-19 pandemic is a global threat that resulted in around 52 million infected people and more than one million deaths globally \cite{dong2020interactive}.
In South America, Chile is among the most severely affected countries, with more than 500 thousand infected people and a death toll that surpassed the 15,000 mark as of November 22nd, 2020. Similarly to other severely affected countries \cite{lai2020effect, haushofer2020interventions, gao2020mobile, chinazzi2020effect, gatto2020spread, jia2020population, tian2020investigation}, Chile implemented Non-Pharmaceutical Interventions (NPIs) such as regional lockdown, stay-at-home orders, and travel restrictions, in an attempt to mitigate the COVID-19 epidemics through reducing individual mobility and promoting social distancing.
In contrast with countries such as China, Italy, and the USA, which implemented NPIs at the national or regional level \cite{bonato2020mobile, cintia2020relationship, lai2020effect, gatto2020spread, tian2020investigation}, Chile's NPIs were implemented at a very localized level, i.e., cities or urban zones (aka \emph{comunas}) \cite{gozzi2020estimating, ferres2020measuring}. Thus, two comunas in the same region may be regulated by different NPIs: whereas one is in lockdown, adjacent ones might have no travel restrictions.
Given the peculiarity of NPIs' spatial scale in Chile, tracking mobility changes is crucial to assess local quarantines' efficacy and measure the effect of mobility reductions on predicting the virus spread \cite{buckee2020aggregated}.

Mobile phone records provide an unprecedented opportunity in tracking human movements \cite{blondel2015survey, barbosa2018human}, allowing for estimating presences and population density \cite{gabrielli2015city, deville2014dynamic, pappalardo2020individual}, mobility patterns \cite{gonzalez2008understanding, pappalardo2015returners, song2010limits, alessandretti2018evidence, barbosa2018human}, flows \cite{hankaew2019interring, balzotti2018understanding, bonnel2018origin}, and socio-economic status \cite{pappalardo2016analytical, eagle2010network, frias2012relation, vscepanovic2015mobile, mao2015quantifying}.
When used correctly and adequately aggregated to preserve privacy \cite{demontoye2018privacy, demontjoye2013unique, pellungrini2017data, pellungrini2020modeling, fiore2019privacy}, mobile phone data represent a crucial tool for supporting public health actions across the phases of the COVID-19 pandemic \cite{oliver2020mobile, buckee2020aggregated}.
Motivated by the potential of mobile phone data in capturing the geographical spread of epidemics \cite{finger2016mobile, tizzoni2014use, wesolowski2012quantifying, bengtsson2015using}, researchers and governments have started to collaborate with mobile network operators to estimate the effectiveness of control measures in several countries \cite{kang2020multiscale, kraemer2020effect, cintia2020relationship, pullano2020population, lai2020effect, liautaud2020fever, badr2020association, coven2020disparities, gozzi2020estimating, bakker2020effect}.

To assess the impact of the NPIs imposed by Chilean authorities in response to the epidemics, we analyse a mobile phone dataset provided by Telef\'onica Chile, which comprises 31 billion eXtended Detail Records (XDRs) and 5.4 million users distributed all over the country covering the period February 26th, 2020 to September 20th, 2020.
An XDR is created every time a user explicitly requests an HTTP address or their device automatically downloads content from the Internet (e.g., emails, text messages), thus describing individual movements in great detail \cite{pappalardo2020individual}.
From the XDRs, we derive three epidemiologically relevant metrics: the Index of Internal Mobility (IM$_{int}$), which quantifies the amount of mobility within each comuna of the country; the Index of External Mobility (IM$_{ext}$), quantifying the mobility between comunas; and the Index of Mobility (IM), which considers any movement, both within and between comunas.
We hence analyse how these metrics change as the COVID-19 epidemics spread out in Chile, highlighting a considerable heterogeneity of response to local quarantines across the country.

The datasets we make available will grow as time goes by and, to the best of our knowledge, are the only ones describing mobility changes and dates of local quarantines in Chile. 
They can be used not only for fighting against the COVID-19 epidemics but will also benefit other research and applications such as emergency response \cite{han2019cities, xu2017collective} and crowd flow prediction \cite{zhang2017deep, xie2020urban, yin2020comprehensive}. The datasets described here are currently used at all levels of the Chilean government.

\section{Methods}
\label{sec:mobile-phone-dimens}
\begin{figure}
    \centering
    \includegraphics[width=1\textwidth]{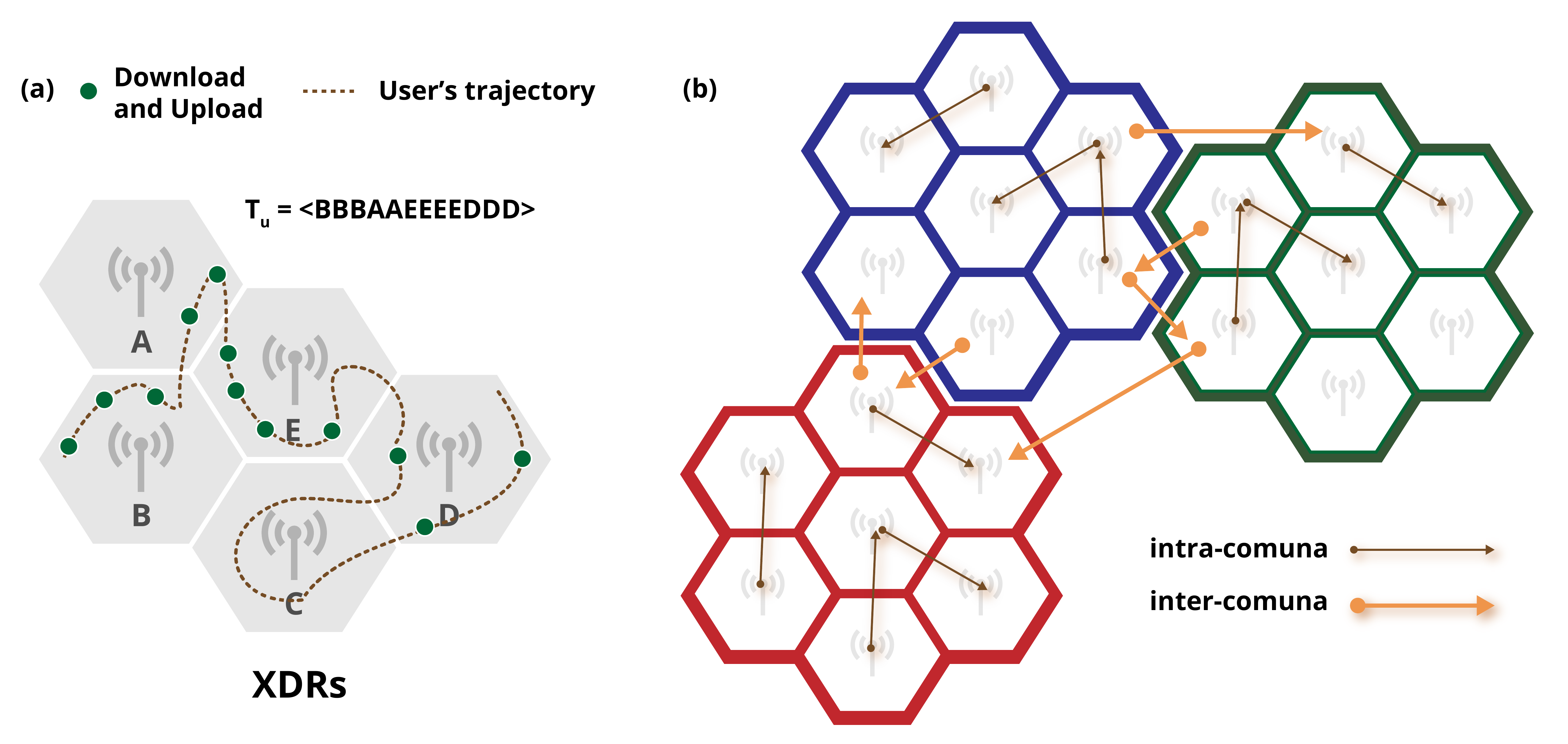}
    \caption{(a) Illustrative example of eXtended Detail Records (XDRs) of a mobile phone user. 
    The hexagons represent mobile phone towers and green dots the positions where the user starts a download/upload operation.
    The dotted line indicates the real movement of the user, from the left to the right. 
    (b) Intra-comuna trips (black arrows) and inter-comuna trips (orange arrows). Hexagons of the same color indicate antennas that fall in the same comuna.
    }
    \label{fig:example_xdrs}
\end{figure}

Mobile phone operators collect several different streams of mobile phones interaction with the cellular network for billing and operational purposes. Among them are the eXtended Detail Records (XDRs), a mixture of human- and device-triggered, either by explicitly requesting an HTTP address or automatically downloading content from the Internet (e.g., emails). 
Formally, an XDR is a tuple $(u,t,A,k)$, in which there is only one antenna $A$ involved, $u$ is the caller's identifier, $t$ is a timestamp of when the record is created, and $k$ is the amount of downloaded information (Figure \ref{fig:example_xdrs}a). 
From the XDRs of the individuals, we define two types of trips. 
Every time a user moves from an antenna to another antenna \textit{within the same comuna}, they generate an intra-comuna trip. 
Every time the user moves from an antenna to an antenna in a different comuna, they generate an inter-comuna trip (Figure \ref{fig:example_xdrs}b). 
For each day and comuna, we construct three indicators of mobility based on the intra- and inter-comuna trips:
\begin{enumerate}
\item $\mbox{IM}_{int}$ (Index of Internal Mobility), the number of intra-comuna trips for that day;
\item $\mbox{IM}_{ext}$ (Index of External Mobility), the number of inter-comuna trips for that day;
\item $\mbox{IM} = \mbox{IM}_{int}+\mbox{IM}_{ext}$ (Index of Mobility).
\end{enumerate}

All the three indices ranges in $[0, \infty)$, where a value of 0 indicates no mobility at all.
We normalize the three indices with respect to the number of users that reside in the comuna, estimated as the total number of unique mobile devices whose home antenna falls in that comuna. 
Each device's home antenna is computed as the antenna in which it has the highest number of XDRs during nighttime (between 7pm and 7am, inclusive) \cite{pappalardo2020individual, vanhoof2018assessing}.
The number of estimated resident users in the comunas is strongly correlated ($R^2=0.96$, slope=4.37, intercept=298.30) with the official population of the comunas as per the official 2017 Chilean Census.

\section{Data Records}

The raw datasets were provided by Telefónica/Movistar Chile, a mobile phone company which possesses between 29-32\% of the Chilean mobile phone market. 
From the raw datasets we construct the three mobility indices described above.
The datasets are released under the CC BY 4.0 License and are publicly available at 
\cite{datacitation}.

Table \ref{tab:records} shows the structure of the dataset describing the mobility indices. 
Each record refers to a comuna in Chile and describes: 
\begin{itemize}
\item the official name of the region (\texttt{region}, type:string);
\item the identifier of the region as per the official 2017 Chilean Census (\texttt{rid}, type:string);
\item the official name of the comuna (\texttt{comuna}, type:string);
 \item the identifier of the comuna as per the official 2017 Chilean Census (\texttt{cid}, type:string)\footnote{All maps and their official identifiers can be downloaded from the National Statistics Office of Chile at \url{https://geoine-ine-chile.opendata.arcgis.com/search?tags=Capas\%20Base}};
\item the area of the comuna in km$^2$ (\texttt{area}, type:float); 
\item the values of IM, IM$_{int}$ and IM$_{ext}$ for that day (type:float);
\item the day the IM, IM$_{int}$ and IM$_{ext}$ values refer to (\texttt{date}, type:date).
\end{itemize}

\begin{table}[ht]\centering
\setlength\tabcolsep{5pt}
\scriptsize
\begin{tabular}{ccccc|cccc}
\toprule
region & rid & comuna & cid & area & IM$_{int}$ & IM$_{ext}$ & IM & date\\
\midrule
Los Ríos & 14 & Valdivia & 14101 & 1018.32 & 6.21 & 0.91 & 7.13 & 2020-02-26\\
Los Ríos & 14 & Valdivia & 14101 & 1018.32 & 6.42 & 0.93 & 7.35 & 2020-02-27\\
Los Ríos & 14 & Valdivia & 14101 & 1018.32 & 6.75 & 1.08 & 7.84 & 2020-02-28\\
Los Ríos & 14 & Valdivia & 14101 & 1018.32 & 6.88 & 1.17 & 8.05 & 2020-02-29\\
Los Ríos & 14 & Valdivia & 14101 & 1018.32 & 5.58 & 1.05 & 6.63 & 2020-03-01\\
\vdots & \vdots & \vdots & \vdots & \vdots & \vdots & \vdots & \vdots & \vdots \\
\bottomrule
\end{tabular}
\caption{Structure of the released dataset.}
\label{tab:records}
\end{table}

Table \ref{tab:quarantines} shows the structure of the quarantines dataset.
Each record refers to a quarantine regulation and describes: 

\begin{itemize}
    \item the identifier of the quarantine regulation (\texttt{qid}, type:integer);
    \item the official name of the comuna (\texttt{comuna}, type:string);
    \item the status of the quarantine, that can be either active or not active (\texttt{status}, type:string);
    \item the coverage of the quarantine, that can be either partial, rural, or complete (\texttt{coverage}, type:string);
    \item the date the quarantine started (\texttt{start}, type:date);
    \item the date the quarantine ended, which is `` - '' if it is still active (\texttt{end}, type:date);
    \item the identifier of the comuna as per the official 2017 Chilean Census (\texttt{cid}, type:string);
    \item the area of the quarantine in m$^2$ (\texttt{area}, type:float);
    \item the perimeter of the quarantine (\texttt{perimeter}, type:float).
\end{itemize}

\begin{table}\centering
\setlength\tabcolsep{4pt}
\scriptsize
\begin{tabular}{ccccccccc}
\toprule
qid	& comuna	& status	& coverage	& start	& end	& cid & area &	perimeter \\
\midrule
4	& El Bosque	& Active	& whole	& 2020-04-16 & -	&	13105	 &	2.06e7 &	1.87e4\\

26	& Quinta Normal	& Active	& whole &	2020-04-23  & - &		13126	&	1.70e7 &	2.12e4\\

38 &	Cerrillos	& Active &	whole	& 2020-05-05 & -	&	13102	& 2.41e7 &	2.52e4\\

42	& Conchalí	& Active & whole &	2020-05-08 &	- &	13104	& 1.59e7 &	1.68e4\\

\vdots & \vdots & \vdots & \vdots & \vdots & \vdots & \vdots & \vdots & \vdots \\
\bottomrule
\end{tabular}
\caption{Structure of the quarantines dataset.}
\label{tab:quarantines}
\end{table}

\section{Technical Validation}
\label{sec:validation}

In our analysis, we consider two periods: the pre-quarantine period, from March 9th to March 15th, 2020, and the quarantine period, from June 22nd to June 28th, 2020. 
Although we have two weeks before March 9th, the transition from February to March marks the start of the Fall school semester in Chile. 
In 2020, March 6th was the start of the semester, so we assume that the ``business as usual'' period would be best represented by the week of March 9th until March 15th. 
March 16th marked the start of NPIs in Chile, with the closure of schools, universities and large public gatherings. 
After that, on March 26th, there was a partial lockdown of seven comunas in the Metropolitan Region. By June 22-28, more than half of the population of the country was under quarantine, and mobility was at 40\% reduction.

\begin{figure}
\centering
    \subfigure[]{\label{fig:PreImnorth}
\includegraphics[width=0.3\textwidth]{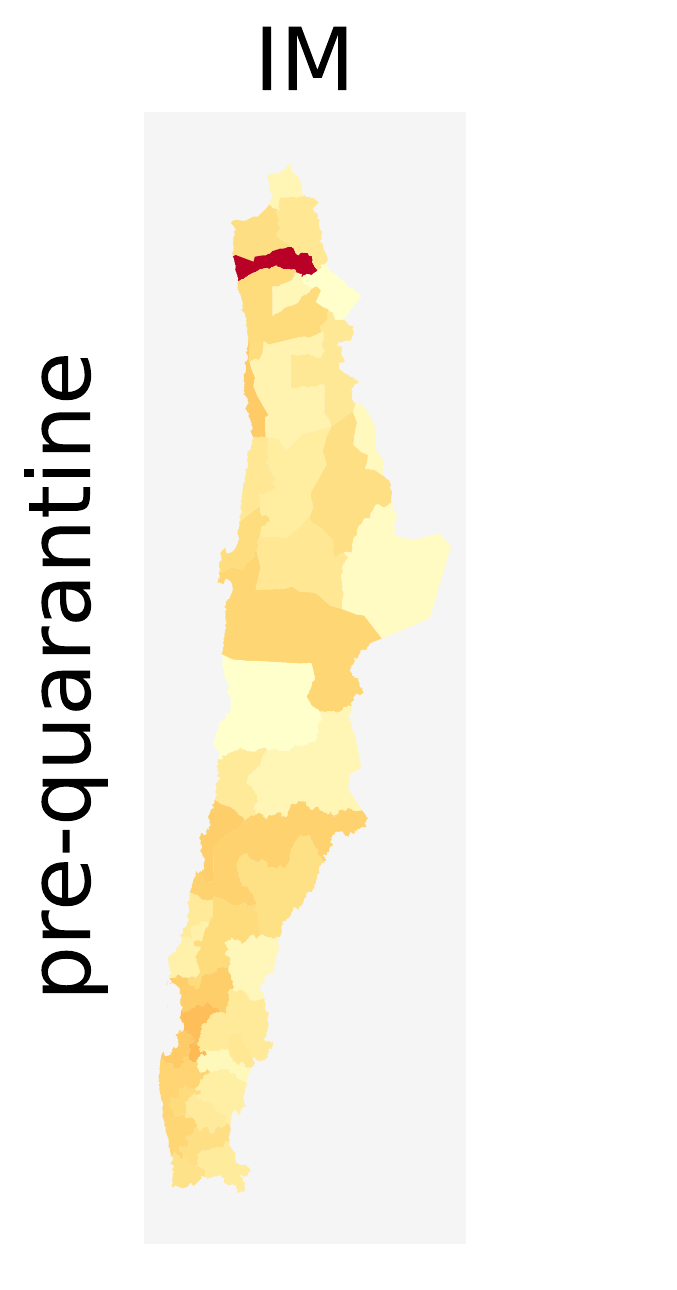}}
    \subfigure[]{\label{fig:PreIMextnorth}
\includegraphics[width=0.3\textwidth]{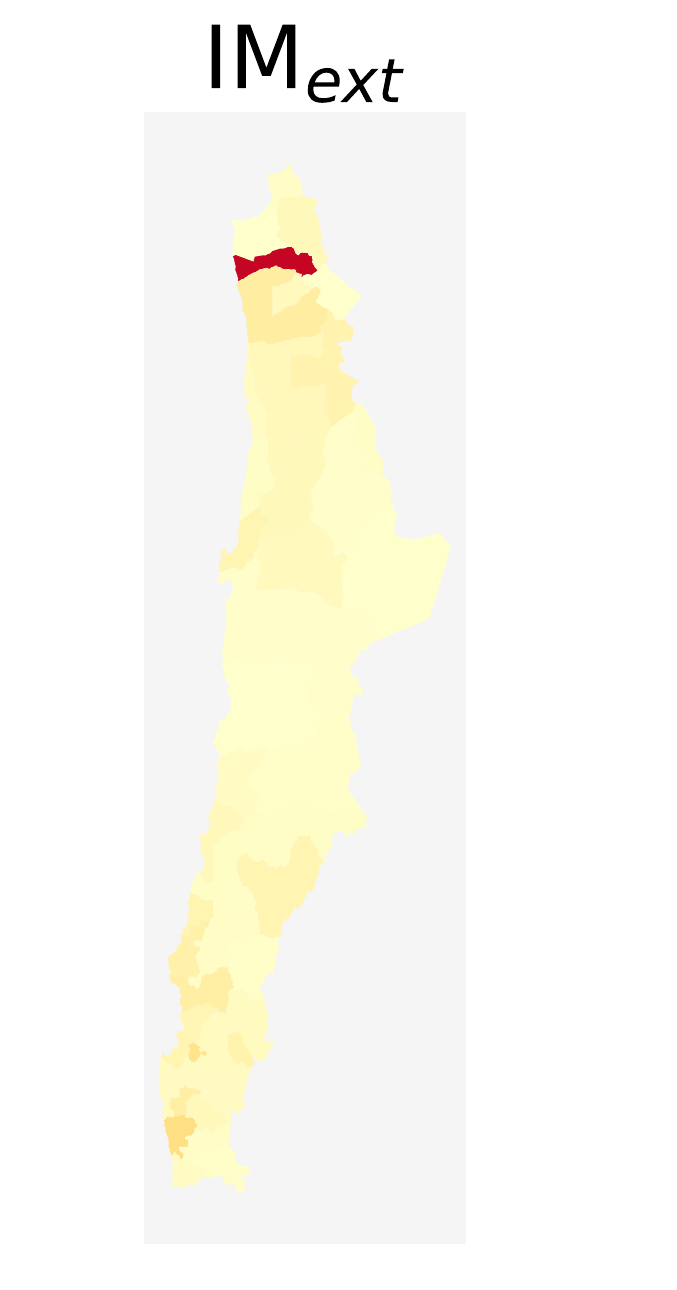}}
    \subfigure[]{\label{fig:PreIMintnorth}
\includegraphics[width=0.3\textwidth]{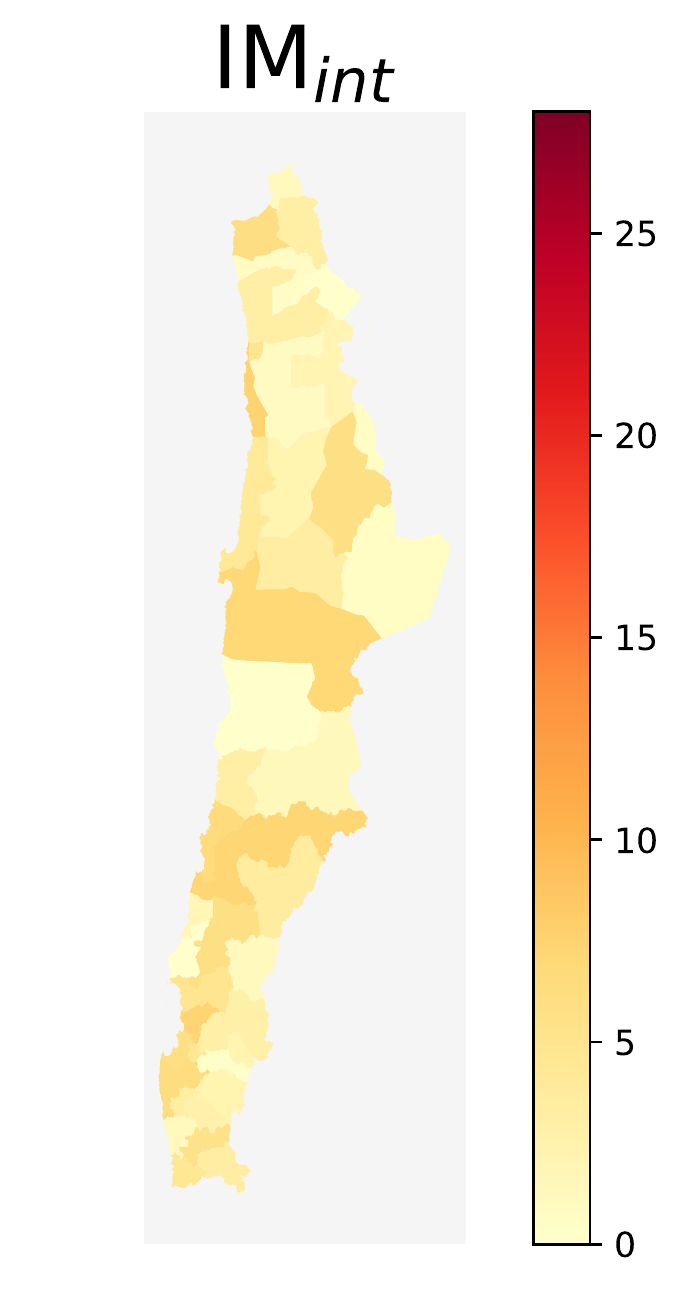}}
\subfigure[]{\label{fig:PanImnorth}
\includegraphics[width=0.3\textwidth]{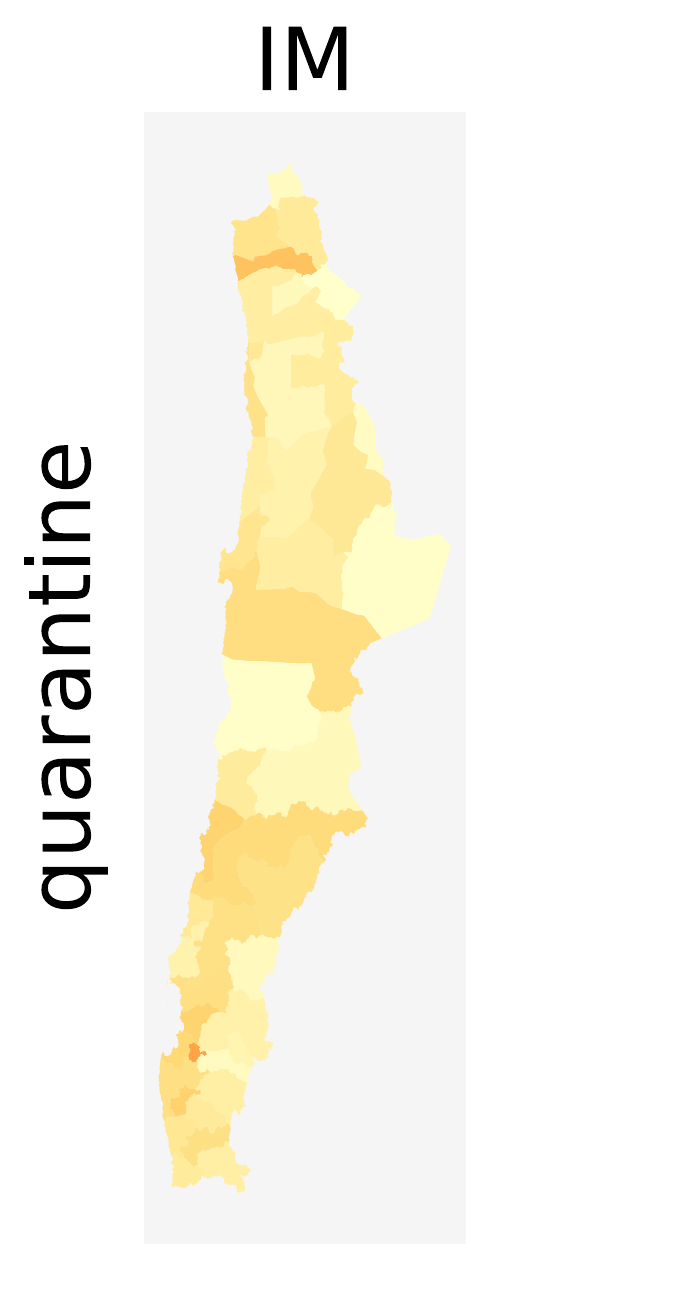}}
\subfigure[]{\label{fig:PanIMextnorth}
\includegraphics[width=0.3\textwidth]{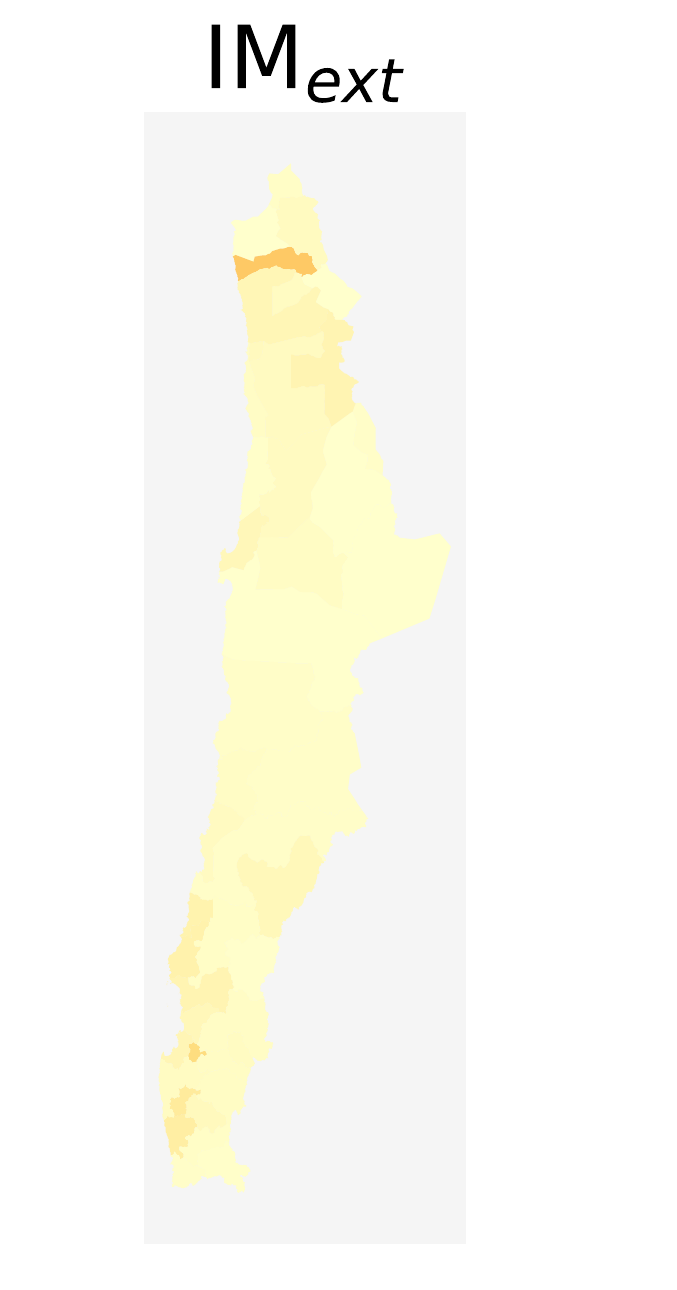}}
\subfigure[]{\label{fig:PanIMintnorth}
\includegraphics[width=0.3\textwidth]{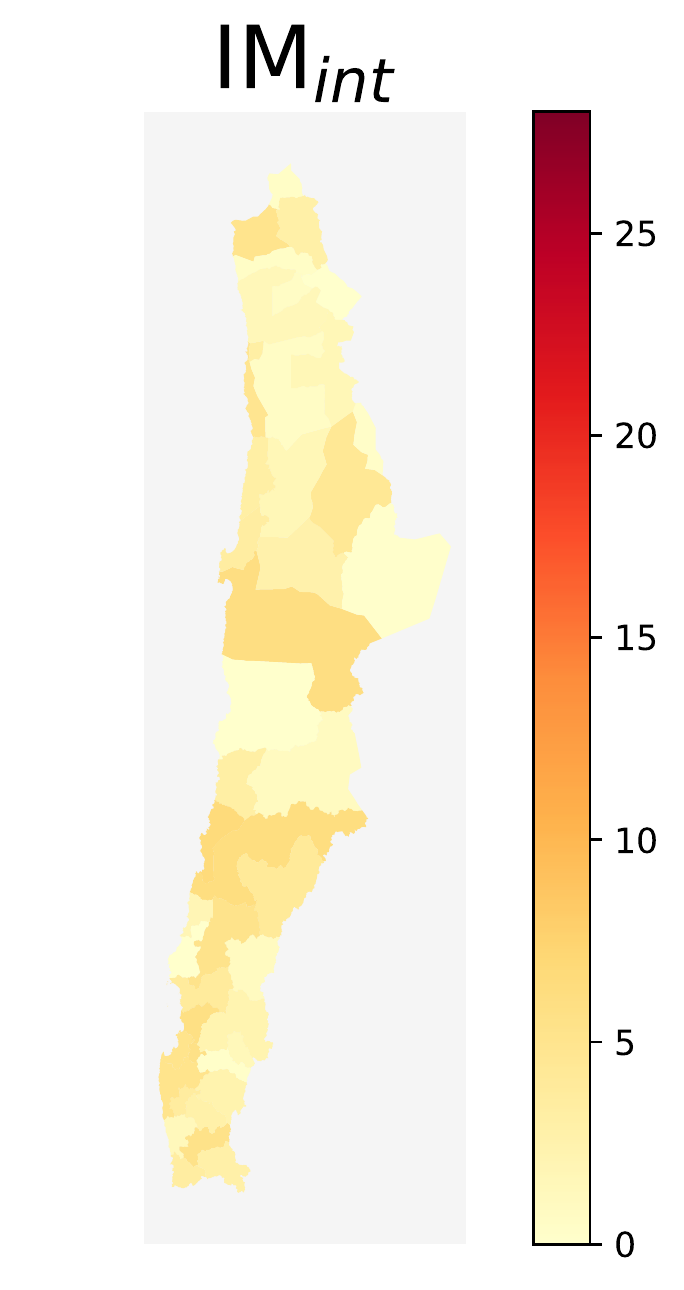}}
\caption{Choropleth maps of IM, IM$_{int}$ and IM$_{ext}$ for the comunas in northern Chile for the pre-quarantine (first row) and the quarantine (second row) periods.
}
\label{fig:Mapsnorth}
\end{figure}

\begin{figure}
\centering
    \subfigure[]{\label{fig:PreImcenter}
\includegraphics[width=0.3\textwidth]{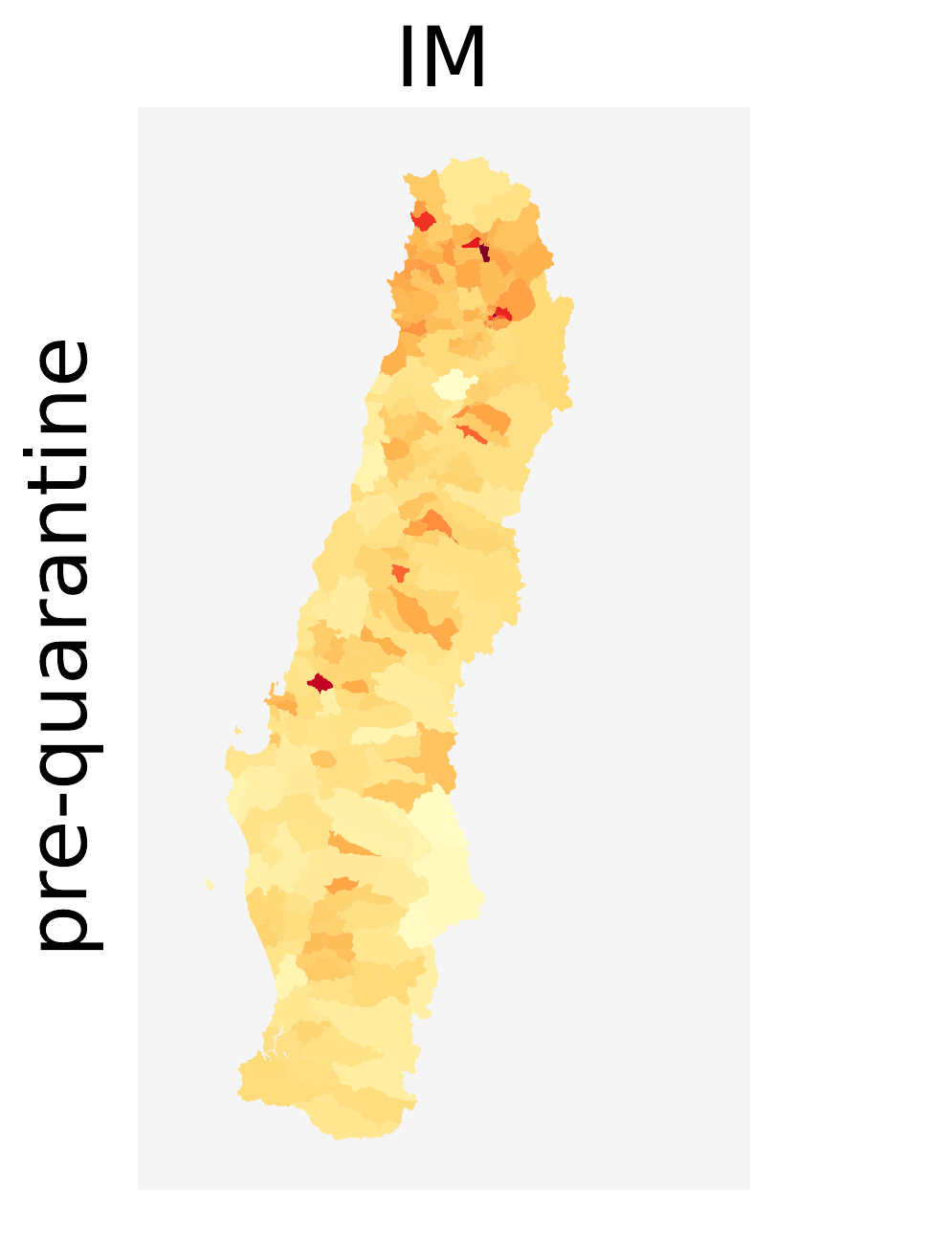}}
    \subfigure[]{\label{fig:PreIMextcenter}
\includegraphics[width=0.3\textwidth]{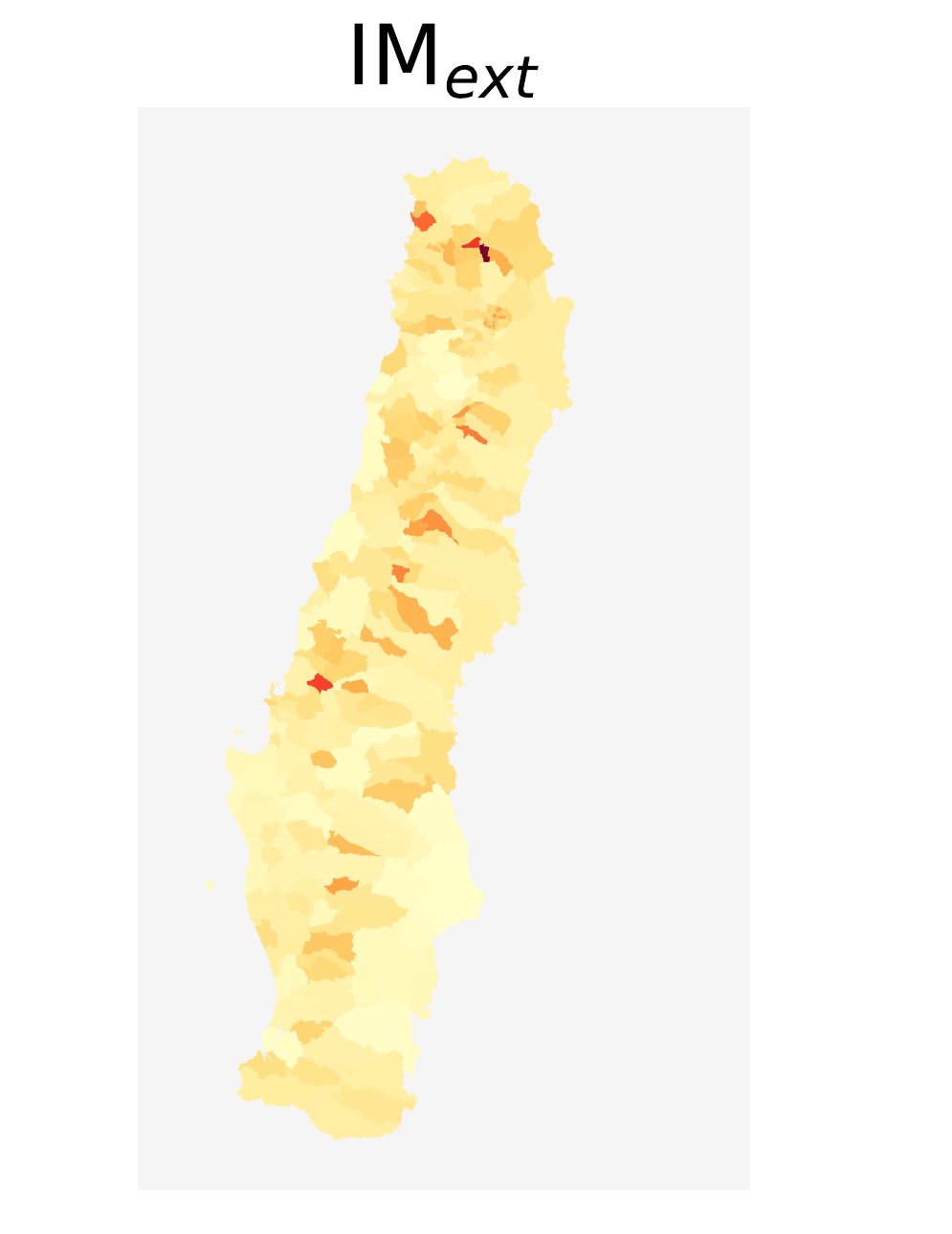}}
    \subfigure[]{\label{fig:PreIMintcenter}
\includegraphics[width=0.3\textwidth]{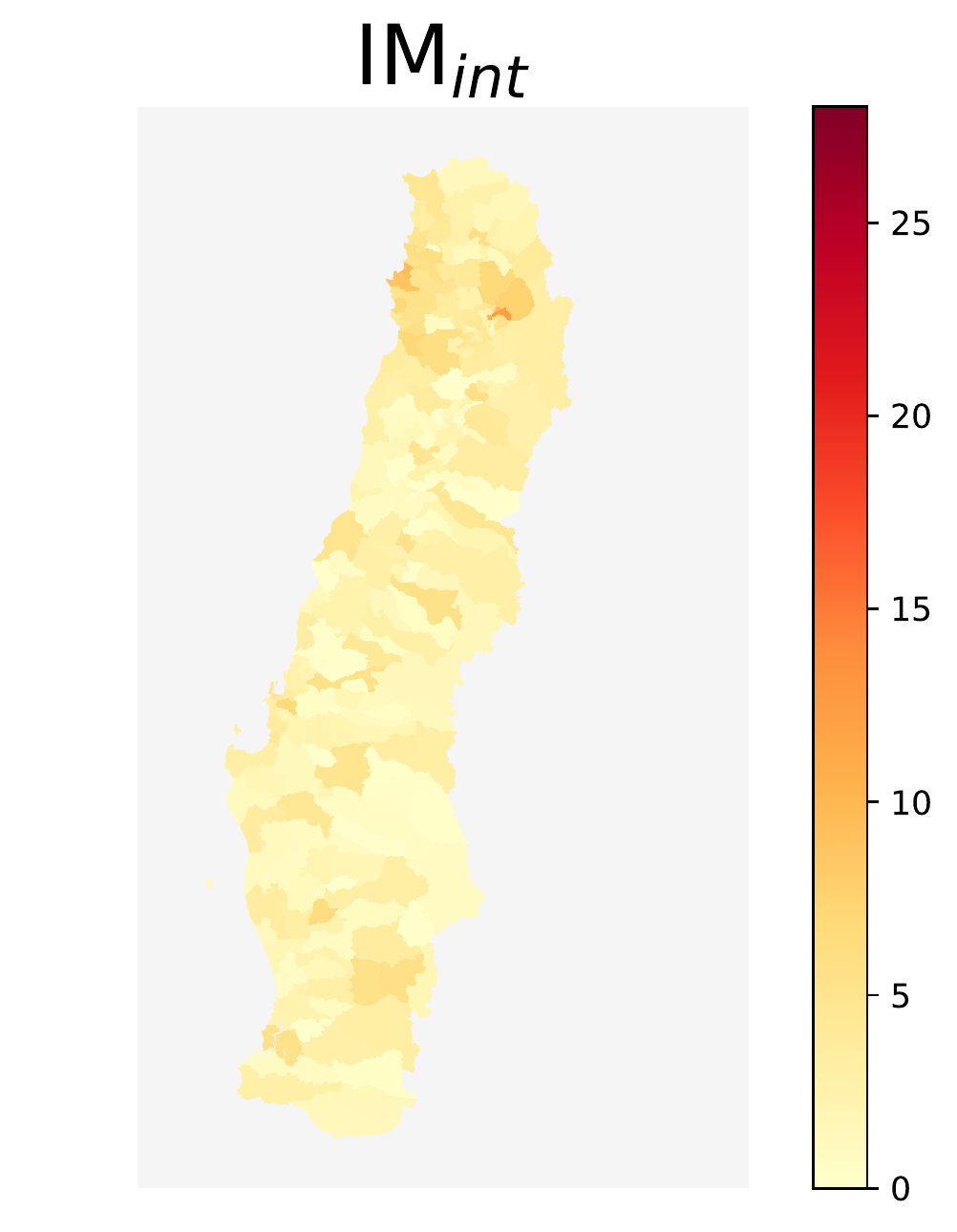}}
\subfigure[]{\label{fig:PanImcenter}
\includegraphics[width=0.3\textwidth]{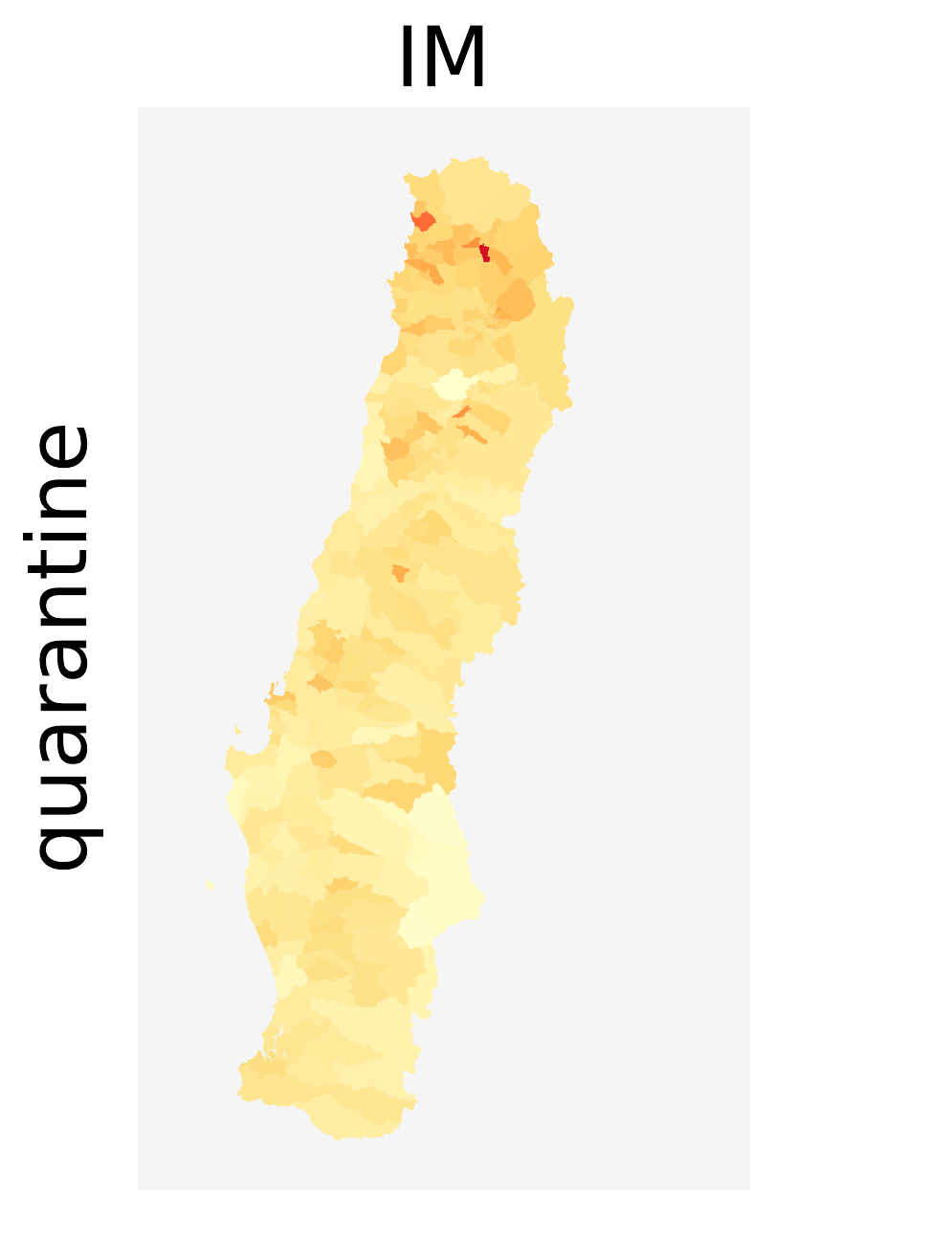}}
\subfigure[]{\label{fig:PanIMextcenter}
\includegraphics[width=0.3\textwidth]{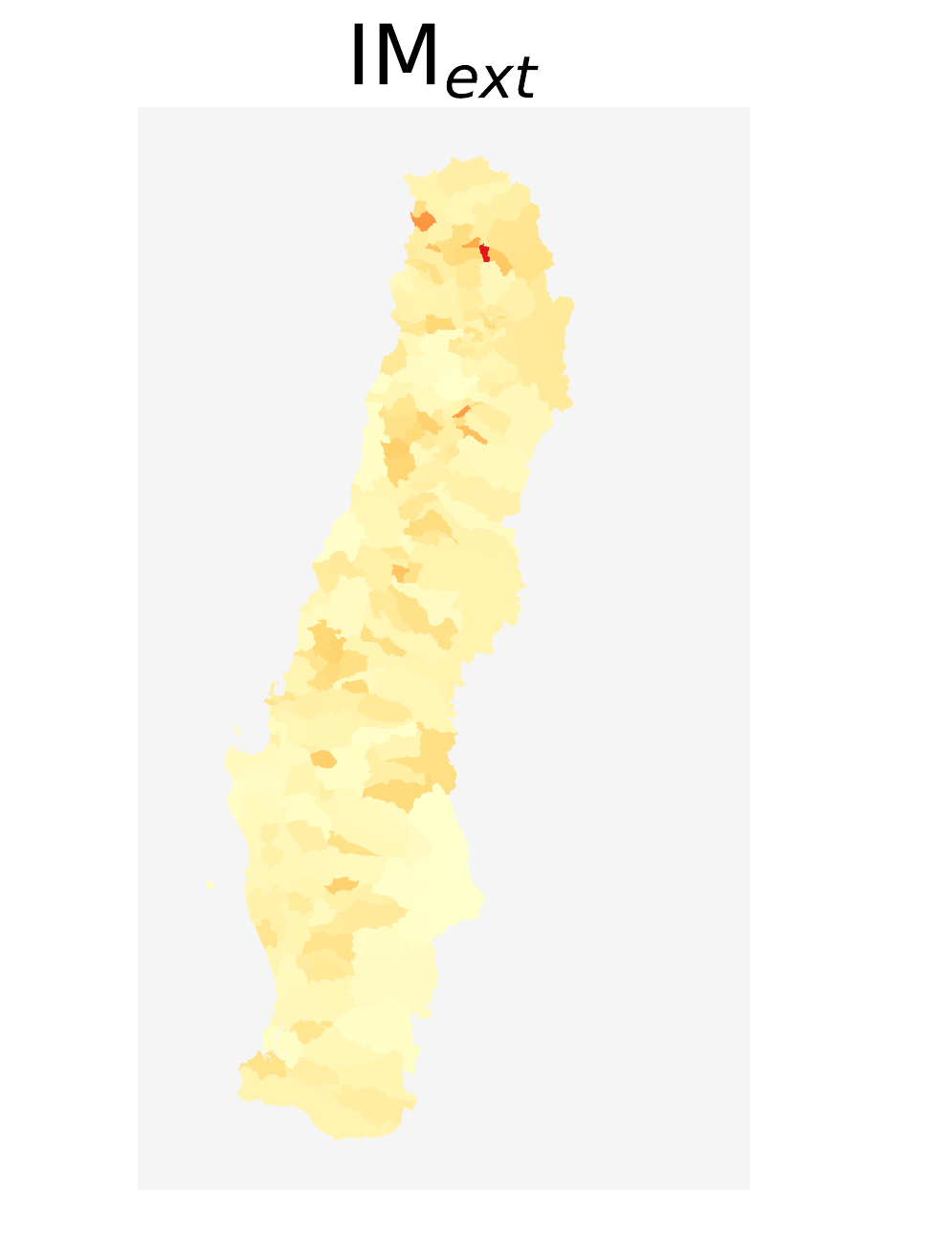}}
\subfigure[]{\label{fig:PanIMintcenter}
\includegraphics[width=0.3\textwidth]{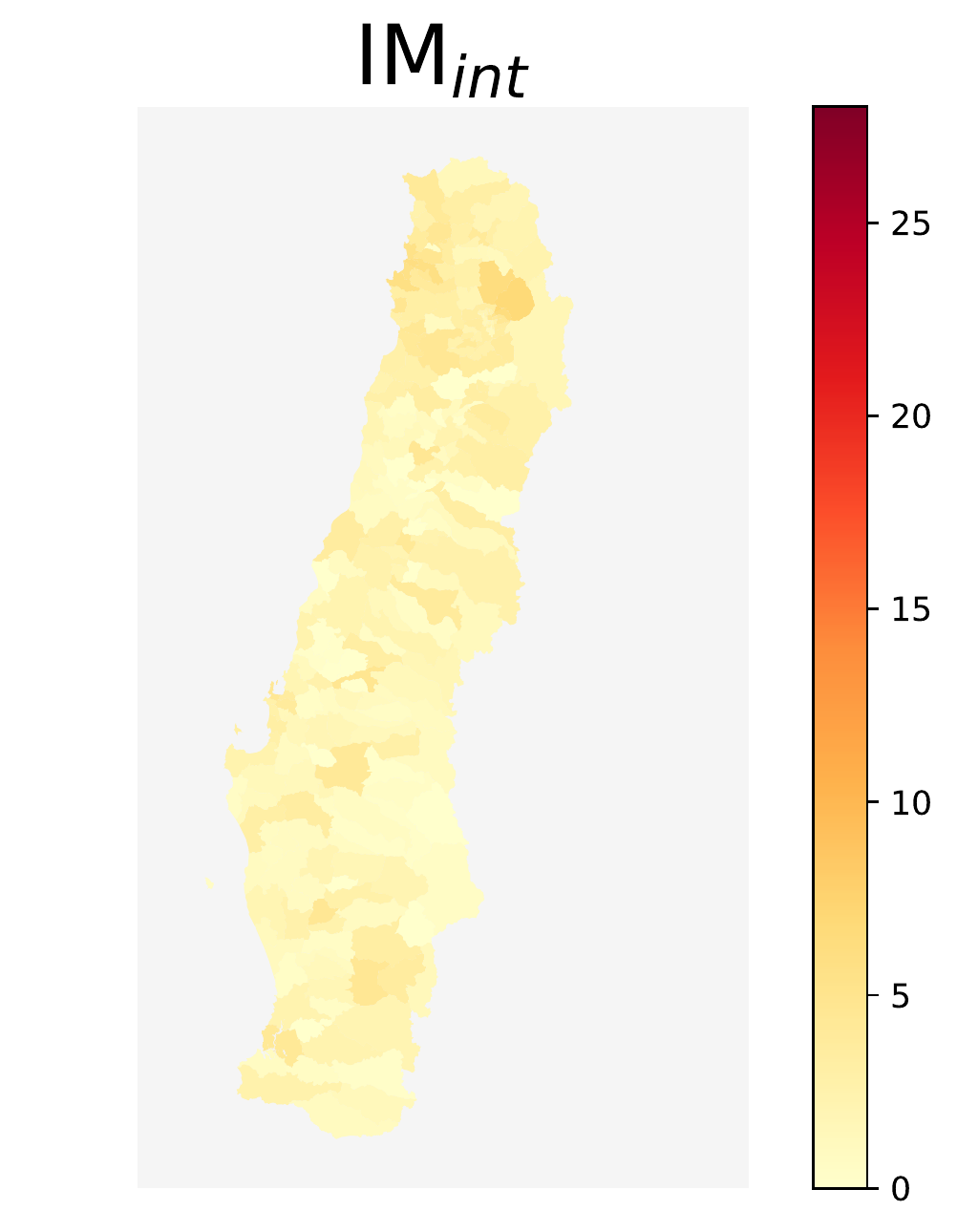}}
\caption{Choropleth maps of IM, IM$_{int}$ and IM$_{ext}$ for the comunas in central Chile for the pre-quarantine (first row) and the quarantine (second row) periods.
}
\label{fig:Mapscenter}
\end{figure}

\begin{figure}
\centering
    \subfigure[]{\label{fig:PreImsouth}
\includegraphics[width=0.3\textwidth]{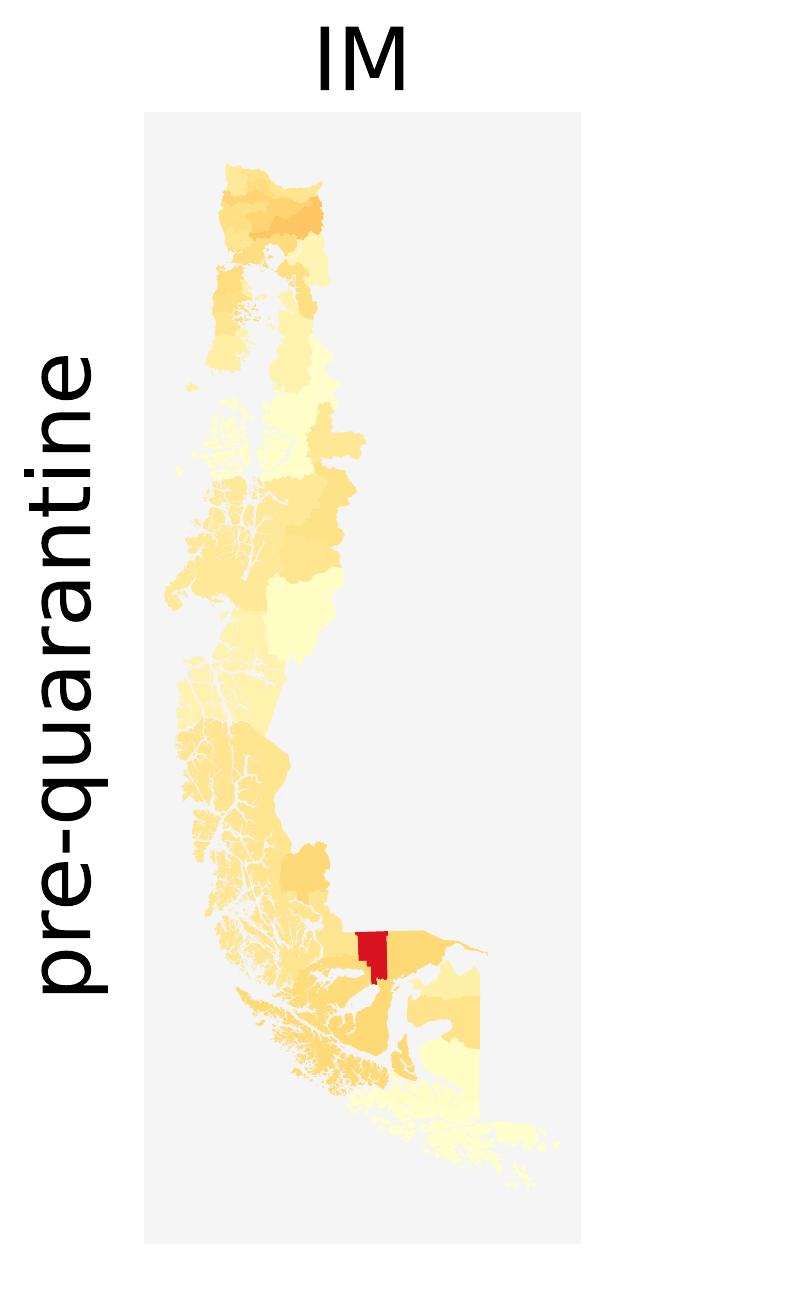}}
    \subfigure[]{\label{fig:PreIMextsouth}
\includegraphics[width=0.3\textwidth]{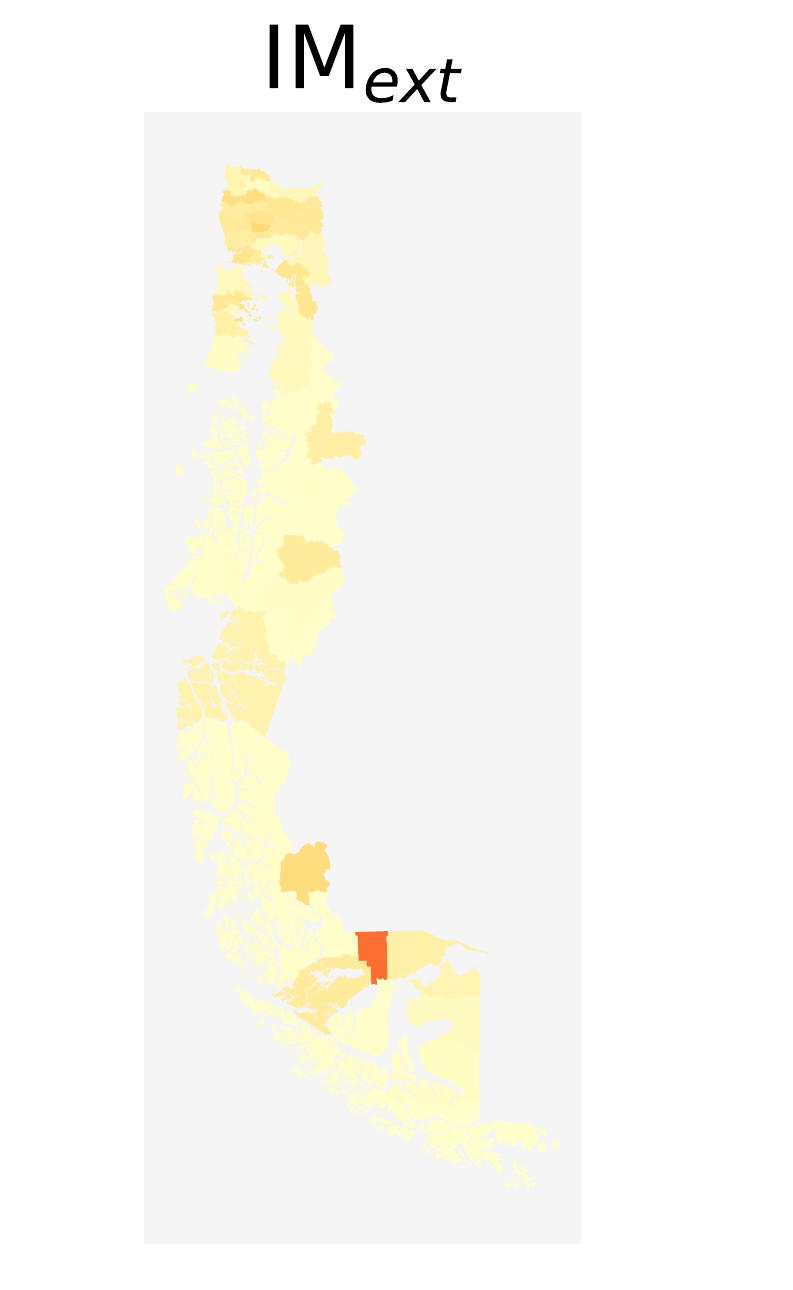}}
    \subfigure[]{\label{fig:PreIMintsouth}
\includegraphics[width=0.3\textwidth]{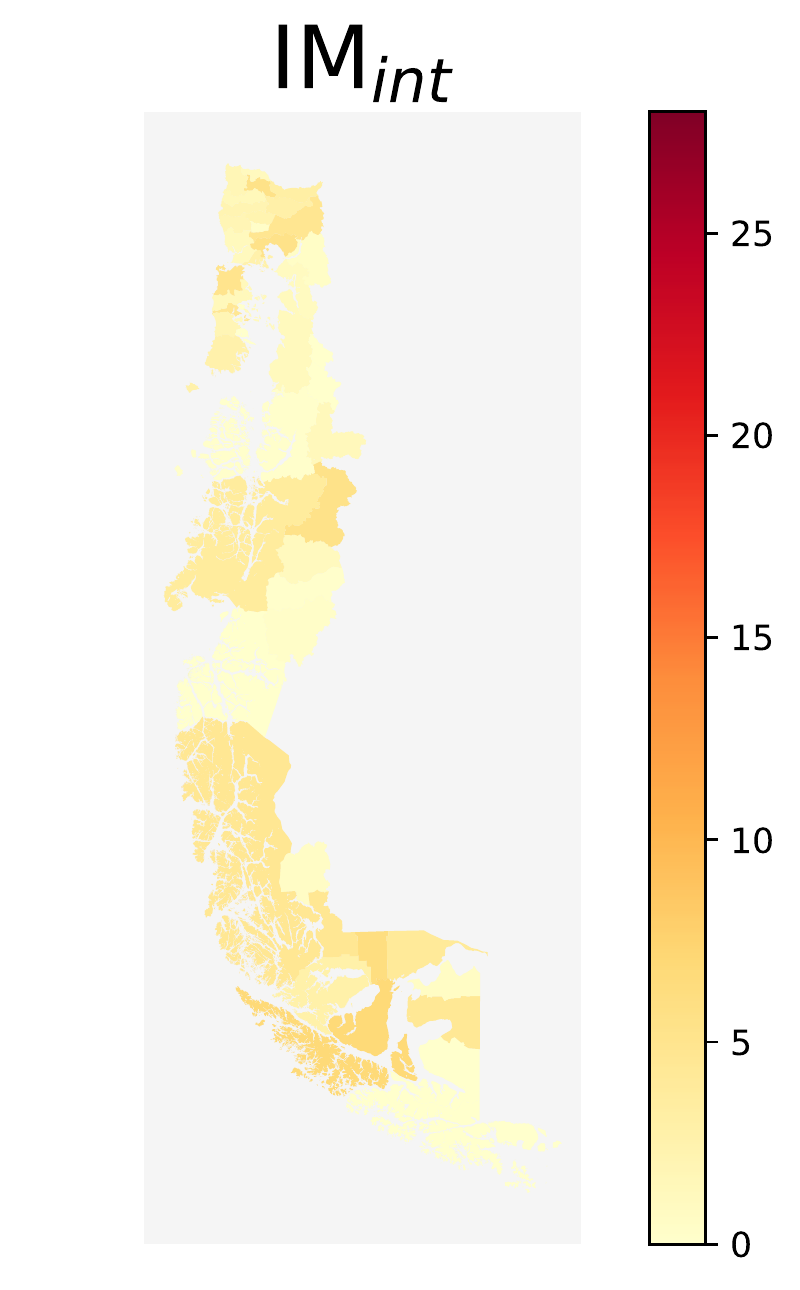}}
\subfigure[]{\label{fig:PanImsouth}
\includegraphics[width=0.3\textwidth]{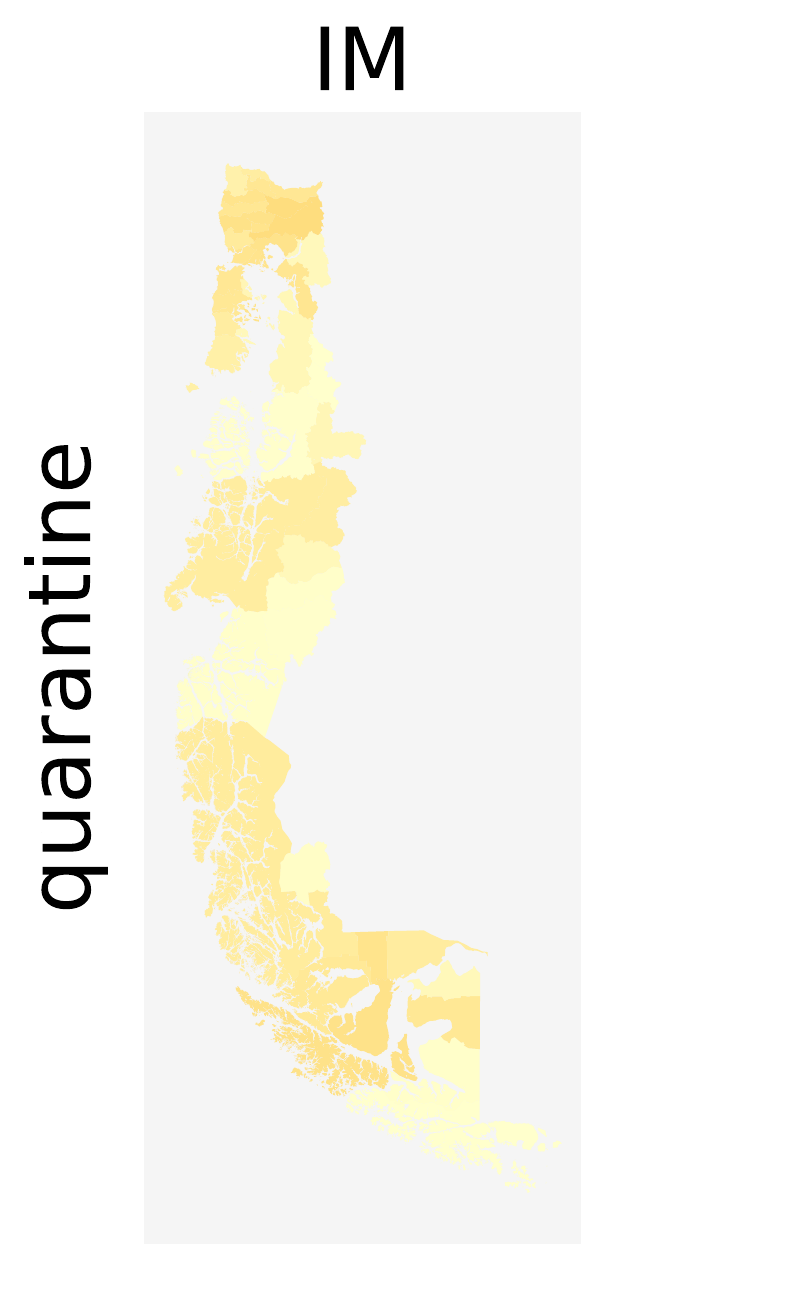}}
\subfigure[]{\label{fig:PanIMextsouth}
\includegraphics[width=0.3\textwidth]{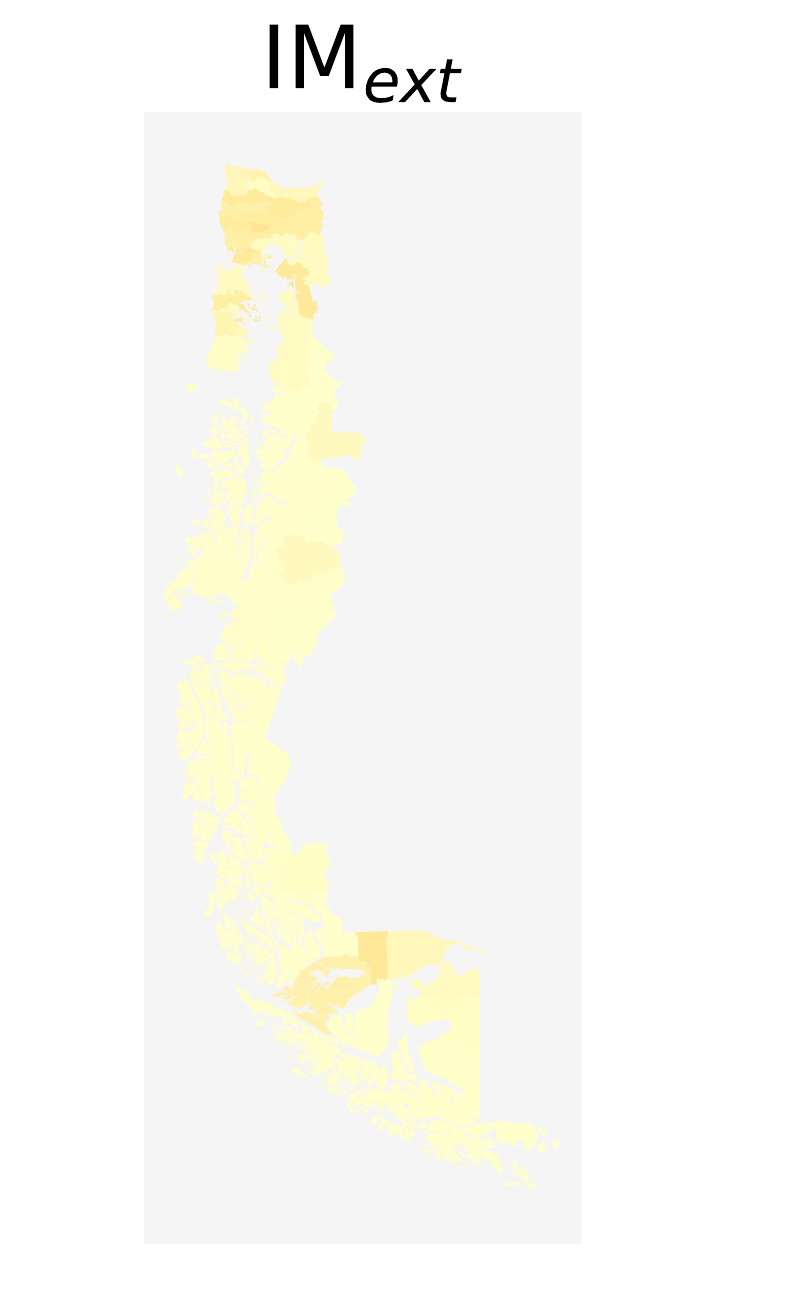}}
\subfigure[]{\label{fig:PanIMintsouth}
\includegraphics[width=0.3\textwidth]{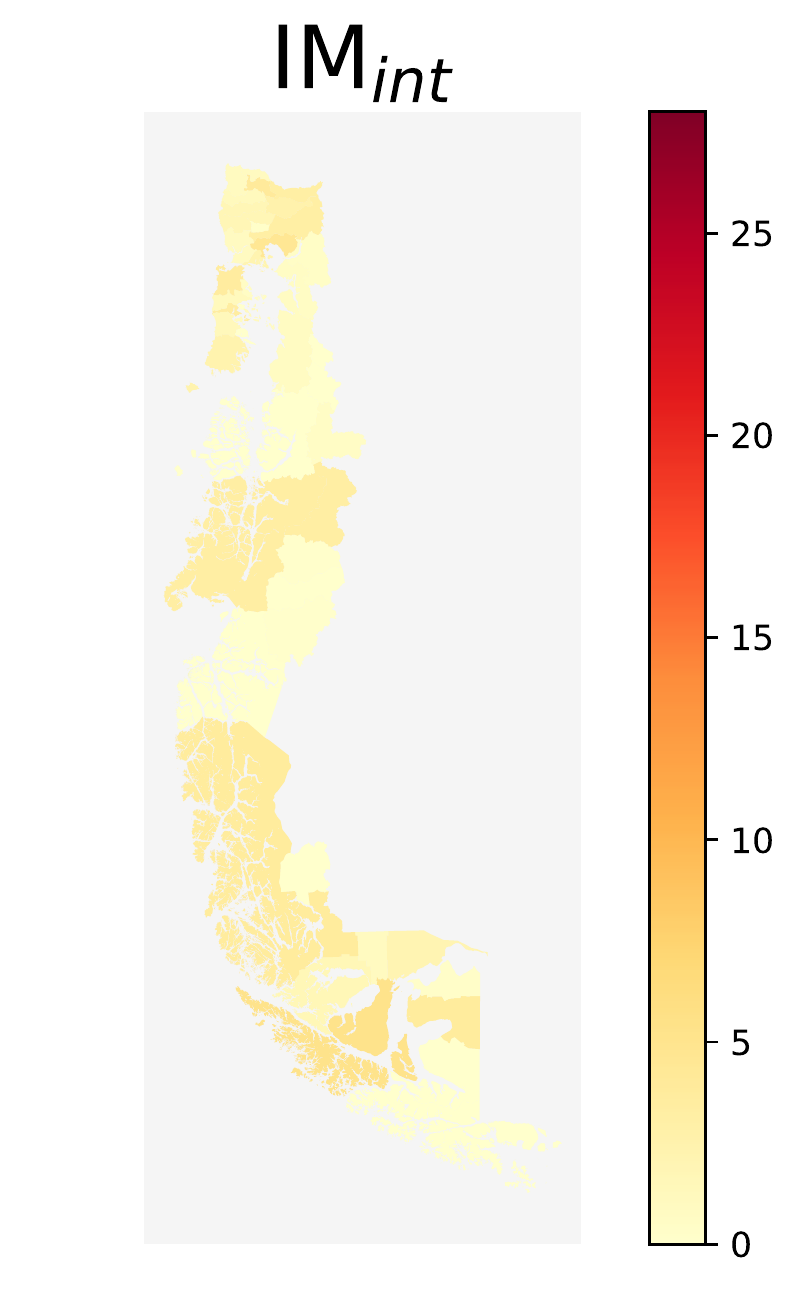}}
\caption{Choropleth maps of IM, IM$_{int}$, and IM$_{ext}$ for the comunas in southern Chile for the pre-quarantine (first row) and the quarantine (second row) periods.
}
\label{fig:Mapssouth}
\end{figure}

\begin{figure}
\centering
    \subfigure[]{\label{fig:PreImSantiagodeChile}
\includegraphics[width=0.3\textwidth]{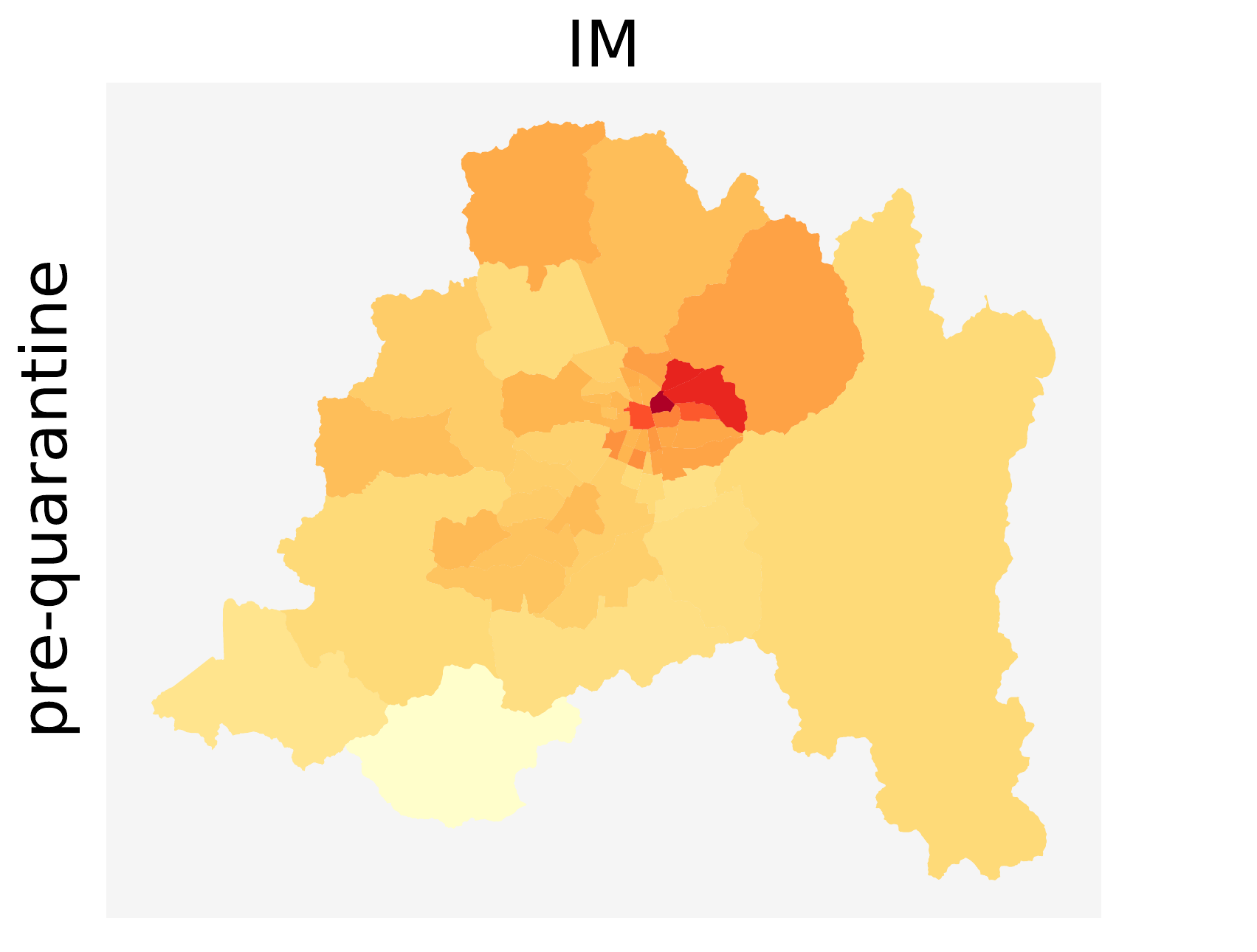}}
    \subfigure[]{\label{fig:PreIMextSantiagodeChile}
\includegraphics[width=0.3\textwidth]{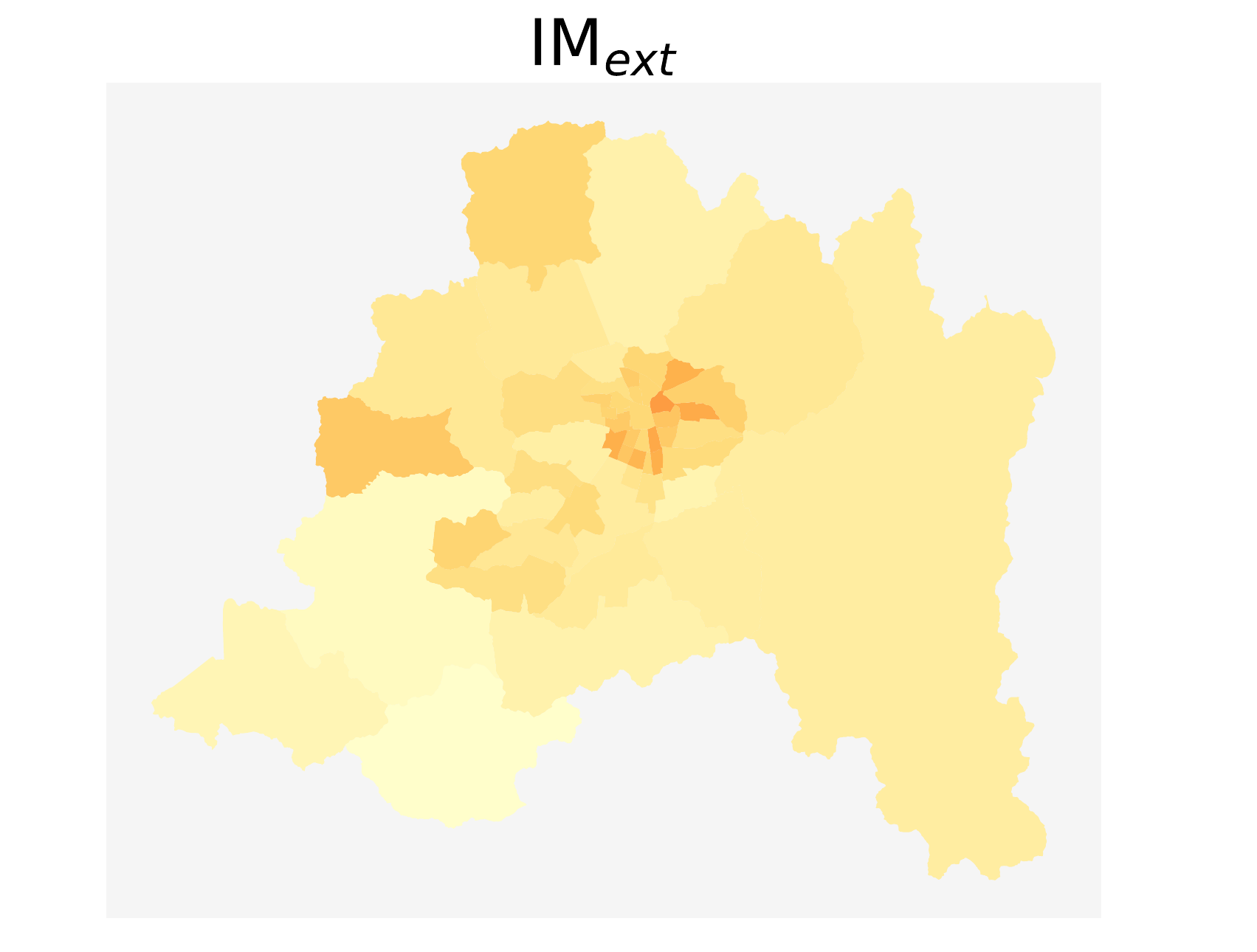}}
    \subfigure[]{\label{fig:PreIMintSantiagodeChile}
\includegraphics[width=0.3\textwidth]{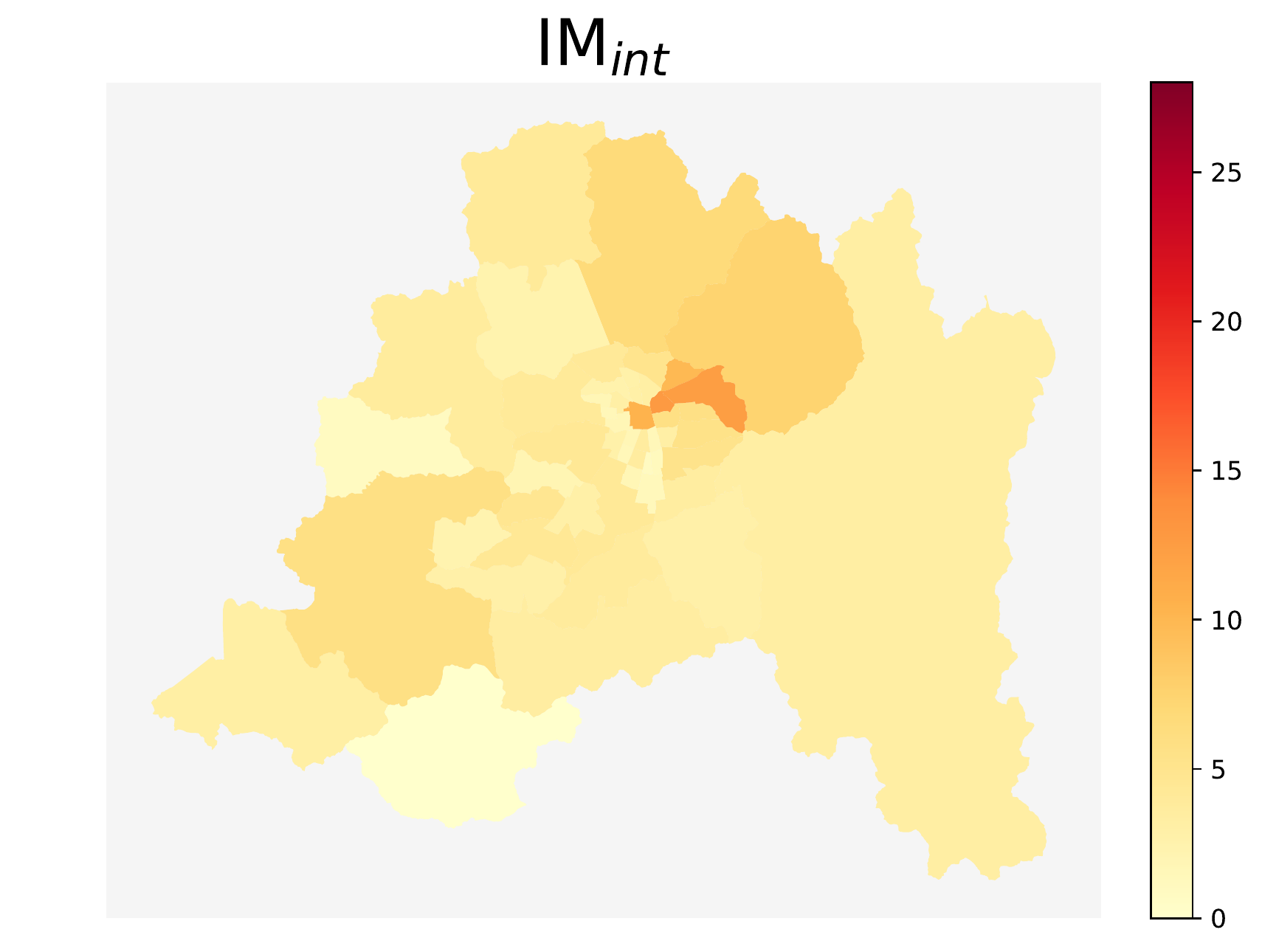}}
\subfigure[]{\label{fig:PanImSantiagodeChile}
\includegraphics[width=0.3\textwidth]{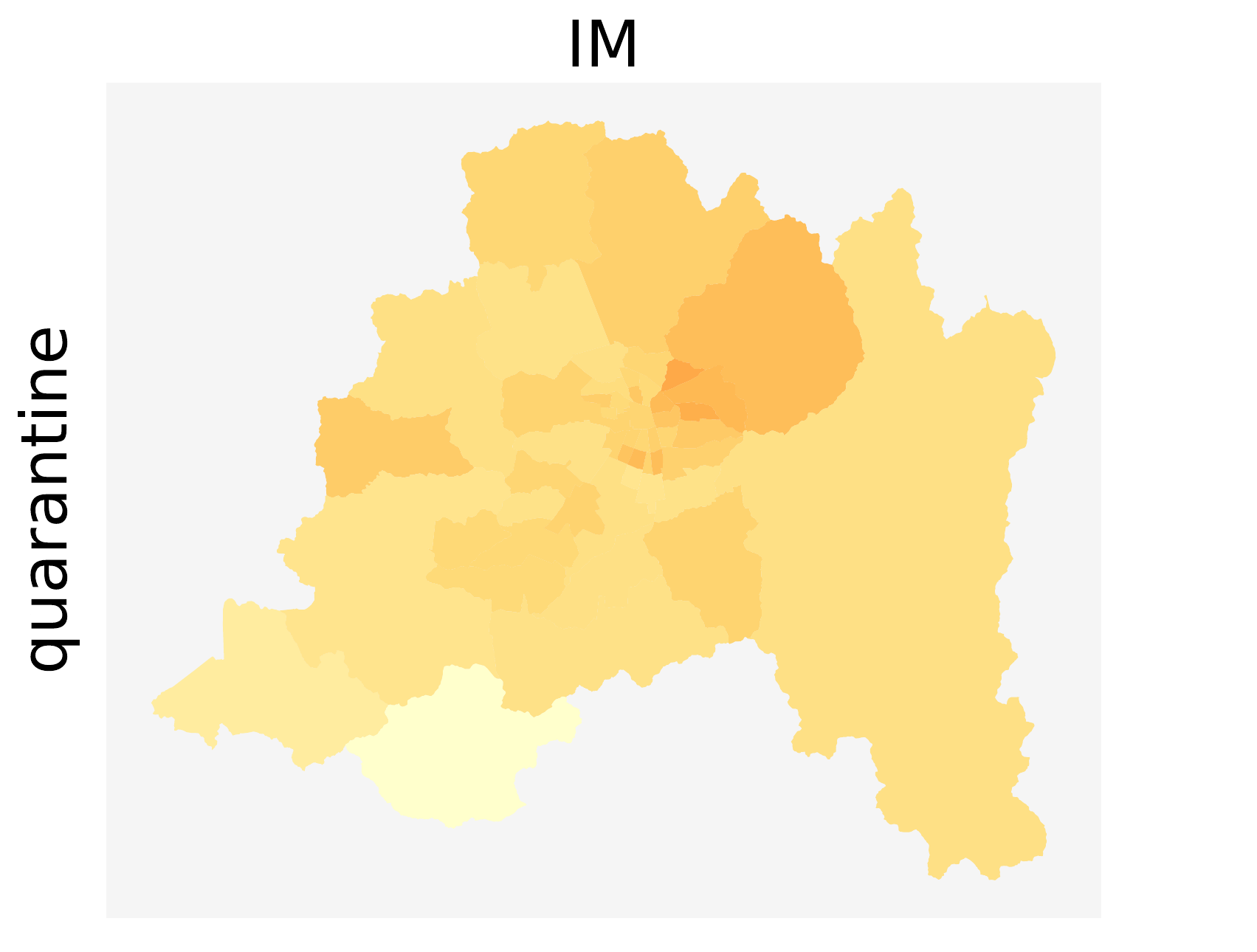}}
\subfigure[]{\label{fig:PanIMextSantiagodeChile}
\includegraphics[width=0.3\textwidth]{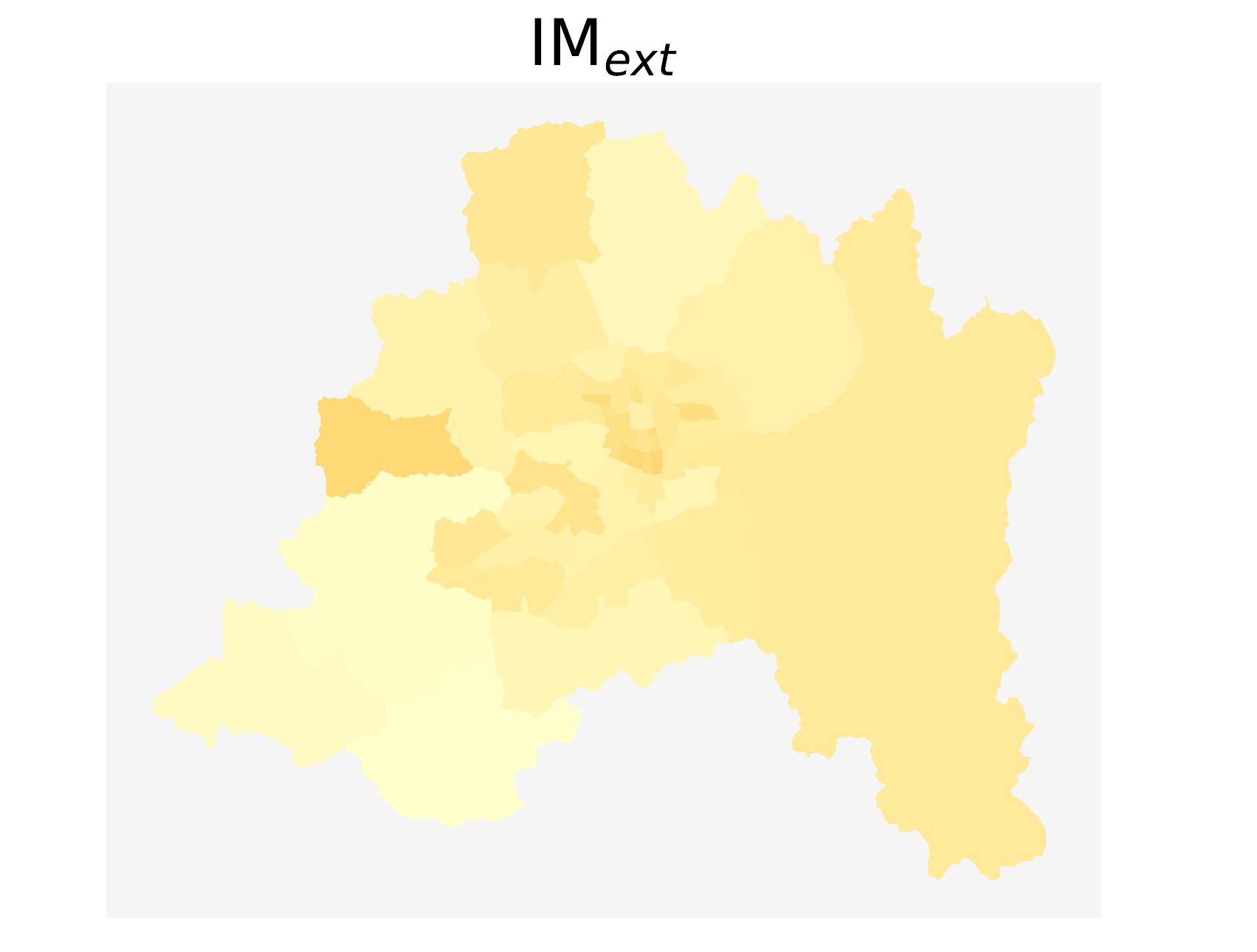}}
\subfigure[]{\label{fig:PanIMintSantiagodeChile}
\includegraphics[width=0.3\textwidth]{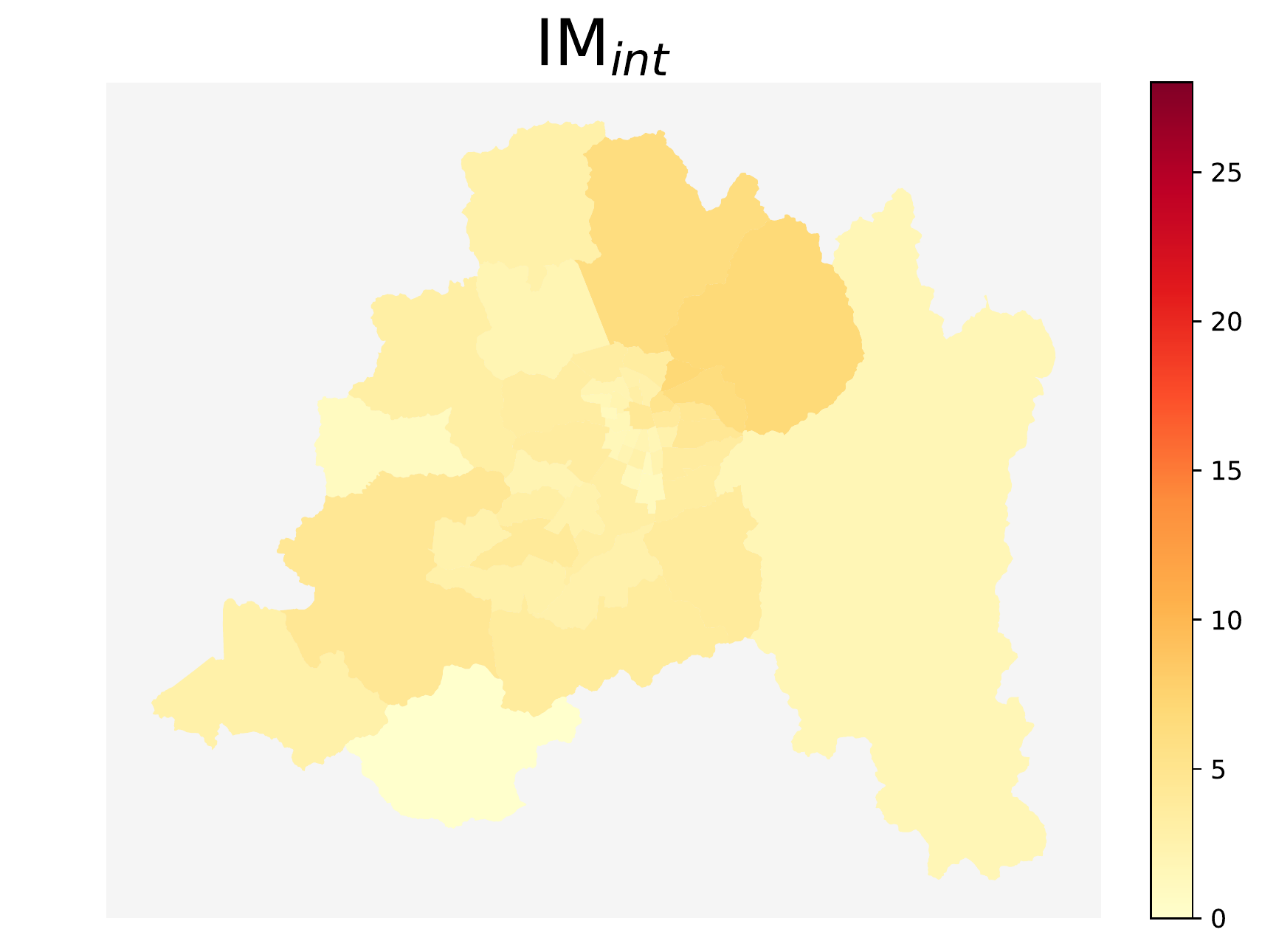}}
\caption{Choropleth maps of IM, IM$_{int}$ and IM$_{ext}$ for the comunas in the metropolitan area of Santiago de Chile for the pre-quarantine (first row) and the quarantine (second row) periods.
}
\label{fig:MapsSantiagodeChile}
\end{figure}

During the pre-quarantine period, comunas with high mobility indices and comunas with low mobility indices coexist.
Geographically, high-mobility comunas are concentrated near urban areas such as the capital Santiago and, in general, in the center of the country (Figures \ref{fig:Mapsnorth}a, \ref{fig:Mapscenter}a, \ref{fig:Mapssouth}a, and \ref{fig:MapsSantiagodeChile}a).
The northern and southern parts of Chile have fewer high-mobility comunas.
The comunas with the highest mobility registered during the pre-quarantine period are located in the regions of Metropolitana de Santiago, Arica y Parinacota, Valparaíso, Ñuble, and Magallanes (Table \ref{tab:rank1}).

\begin{table}[ht]\centering
\begin{tabular}{@{}cccccc@{}}
\multicolumn{6}{c}{\large Pre-quarantine period}\\
\toprule
 & Comuna        & Region                    & IM    & $\mbox{IM}_{ext}$ & $\mbox{IM}_{int}$ \\ \midrule
1    & Rinconada     & Valparaíso                & 30.37 & 27.96   & 2.42    \\
2    & Providencia   & Metropolitana de Santiago & 25.29 & 12.58   & 12.71   \\
3    & Camarones     & Arica y Parinacota        & 24.62 & 23.77   & 0.85    \\
4    & Ranquil       & Ñuble                     & 23.87 & 18.33   & 5.54    \\
5    & Laguna Blanca & Magallanes                & 21.92 & 15.75   & 6.18    \\
6    & Panquehue     & Valparaíso                & 20.93 & 19.02   & 1.90    \\
7    & Vitacura      & Metropolitana de Santiago & 20.40 & 10.54   & 9.86    \\
8    & Las Condes    & Metropolitana de Santiago & 20.22 & 7.79    & 12.42   \\
9    & Zapallar      & Valparaíso                & 19.26 & 15.98   & 3.28    \\
10   & Santiago      & Metropolitana de Santiago & 17.44 & 6.97    & 10.48   \\ \bottomrule
\end{tabular}
\caption{The ten comunas with the highest average value of the IM index computed between March 9th and March 15th, 2020.}
\label{tab:rank1}
\end{table}

\begin{figure}
\centering
    \subfigure[]{\label{fig:tsIM}
\includegraphics[width=1\textwidth]{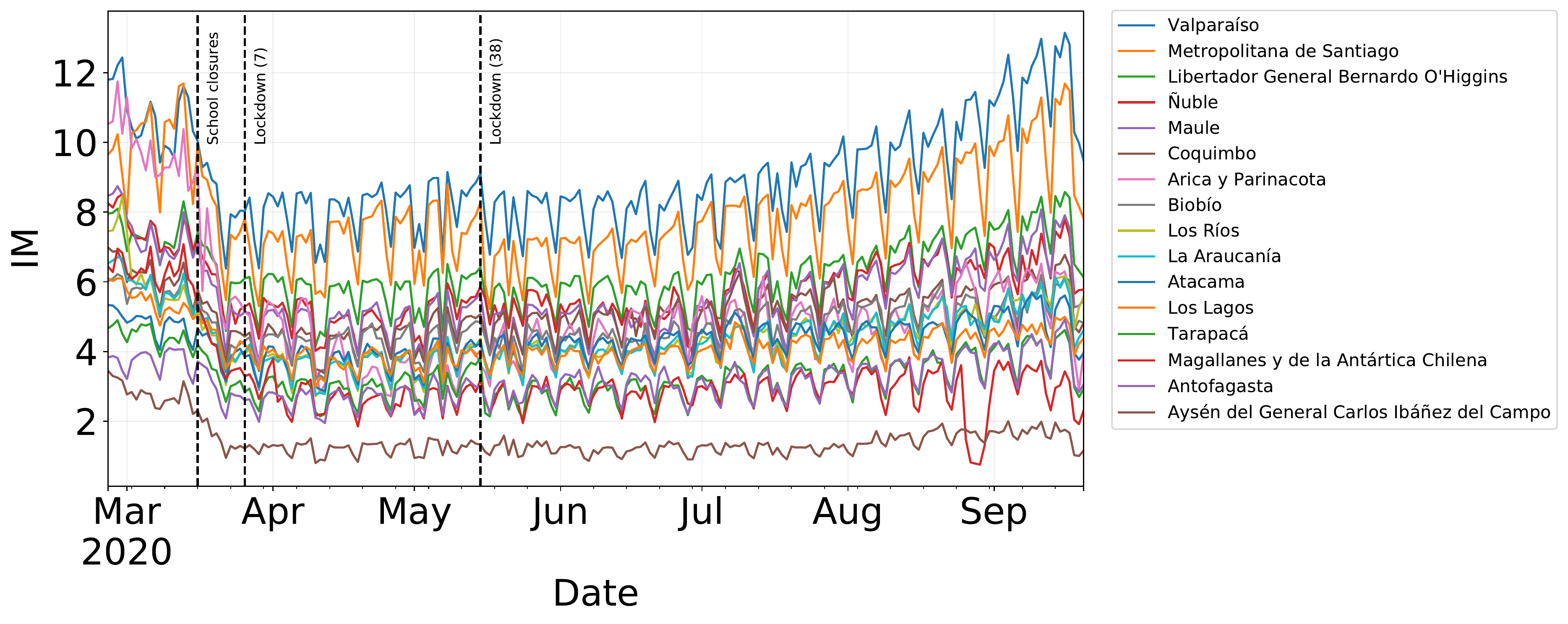}}
\hspace{2mm}
    \subfigure[]{\label{fig:tsIM_ext}
\includegraphics[width=1\textwidth]{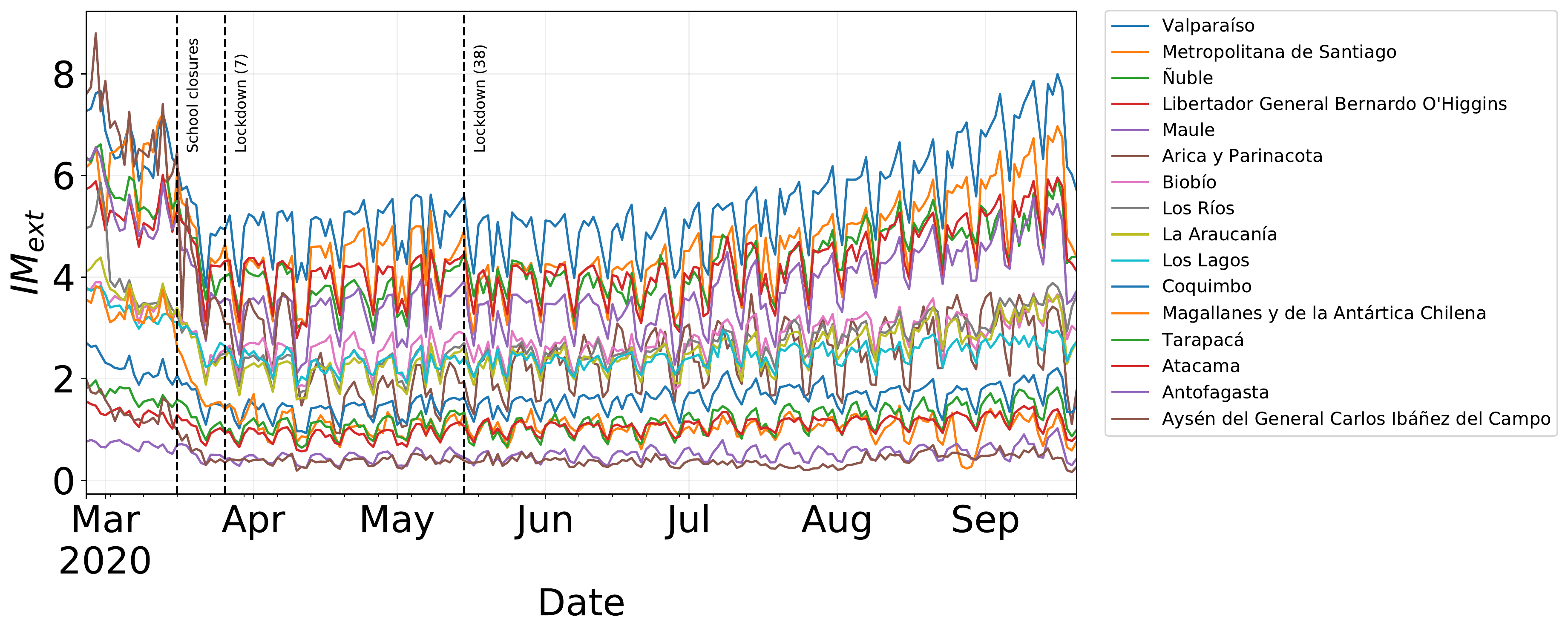}}
    \subfigure[]{\label{fig:tsIM_int}
\includegraphics[width=1\textwidth]{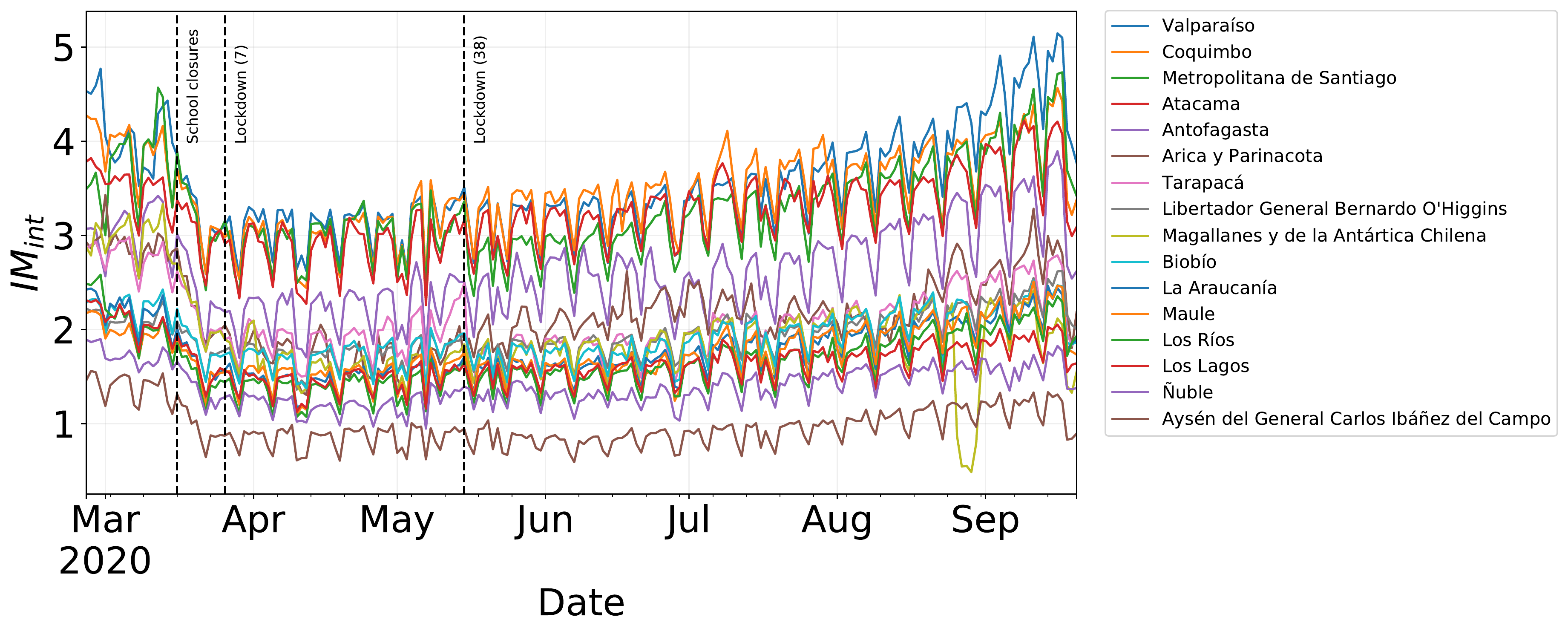}}
\caption{Evolution of IM (a), $\mbox{IM}_{ext}$ (b) and $\mbox{IM}_{int}$ (c) from March to September 2020 for the 16 regions in Chile.
The curves are sorted in descending order respect to the relative index of mobility of the corresponding comuna.
The vertical lines denote important dates regarding NPIs in Chile;
the number in parentheses indicates the number of comunas subject to that restriction.}
\label{fig:timeseriesIM}
\end{figure}

\begin{figure}
    \centering
    \includegraphics[width=1\textwidth]{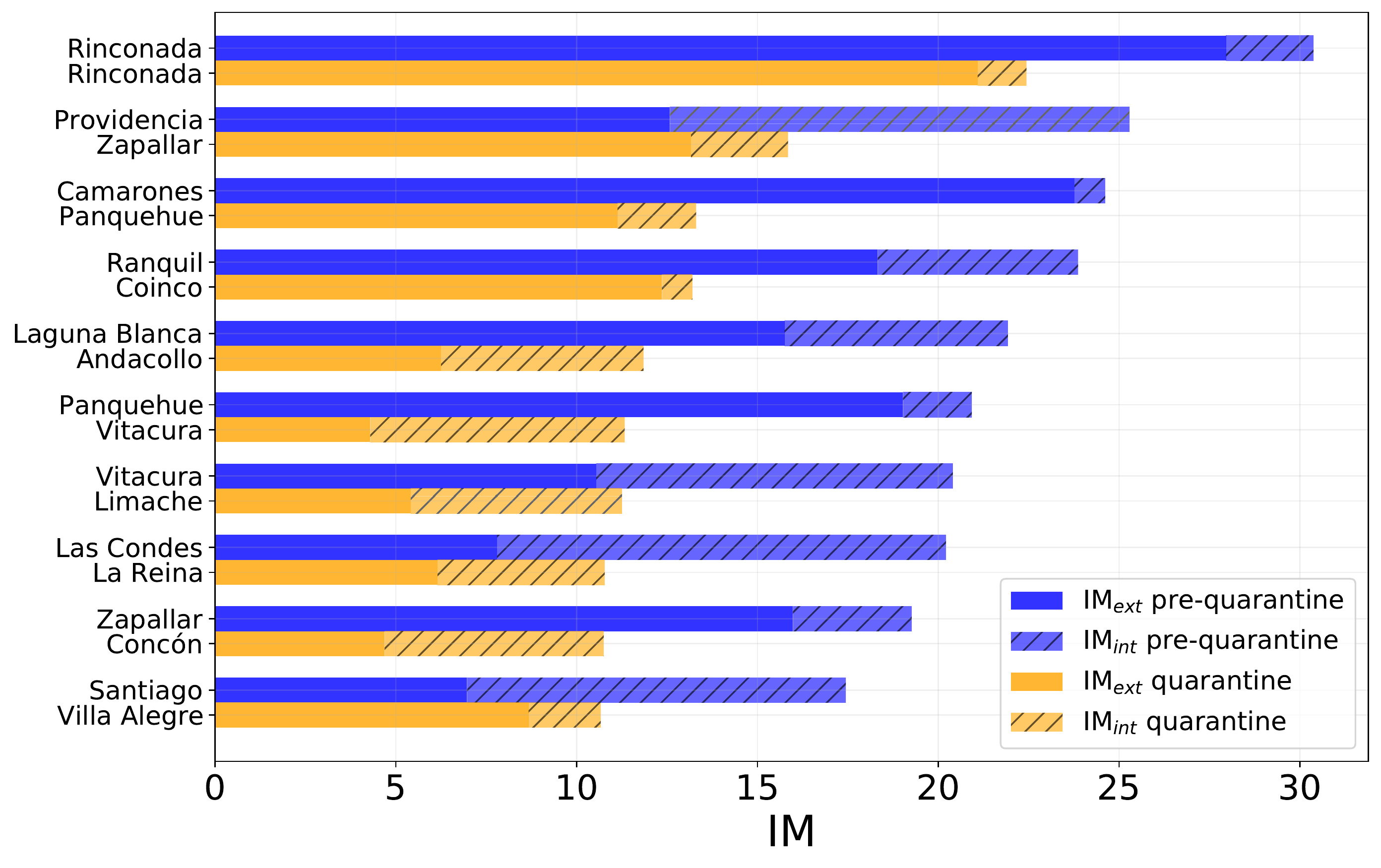}
    \caption{
    Values of IM, IM$_{int}$, and IM$_{ext}$ of the comunas in the top 10 ranking computed for the pre-quarantine and quarantine period.
    The coupled bars represent comunas corresponding to the same position in the rank.
    }
    \label{fig:barplot_ranks}
\end{figure}

The top-ten comunas with the highest mobility indices change during the quarantine period, except for Rinconada in the region of Valparaíso (Table \ref{tab:rank2}), mirroring the different degree of reduction in human mobility in the Chilean regions (Figure \ref{fig:barplot_ranks}).
All regions show a reduction in all three mobility indices during the quarantine period, albeit with different intensities (Figure \ref{fig:timeseriesIM}). 
At the comuna level, high-mobility comunas are rare and clustered near the large urban areas located in central Chile (Figures \ref{fig:Mapsnorth}, \ref{fig:Mapscenter}, \ref{fig:Mapssouth}, and \ref{fig:MapsSantiagodeChile}).

\begin{figure}[!h]
\centering
\subfigure[]{\label{fig:avg_IM}\includegraphics[width=0.45\textwidth]{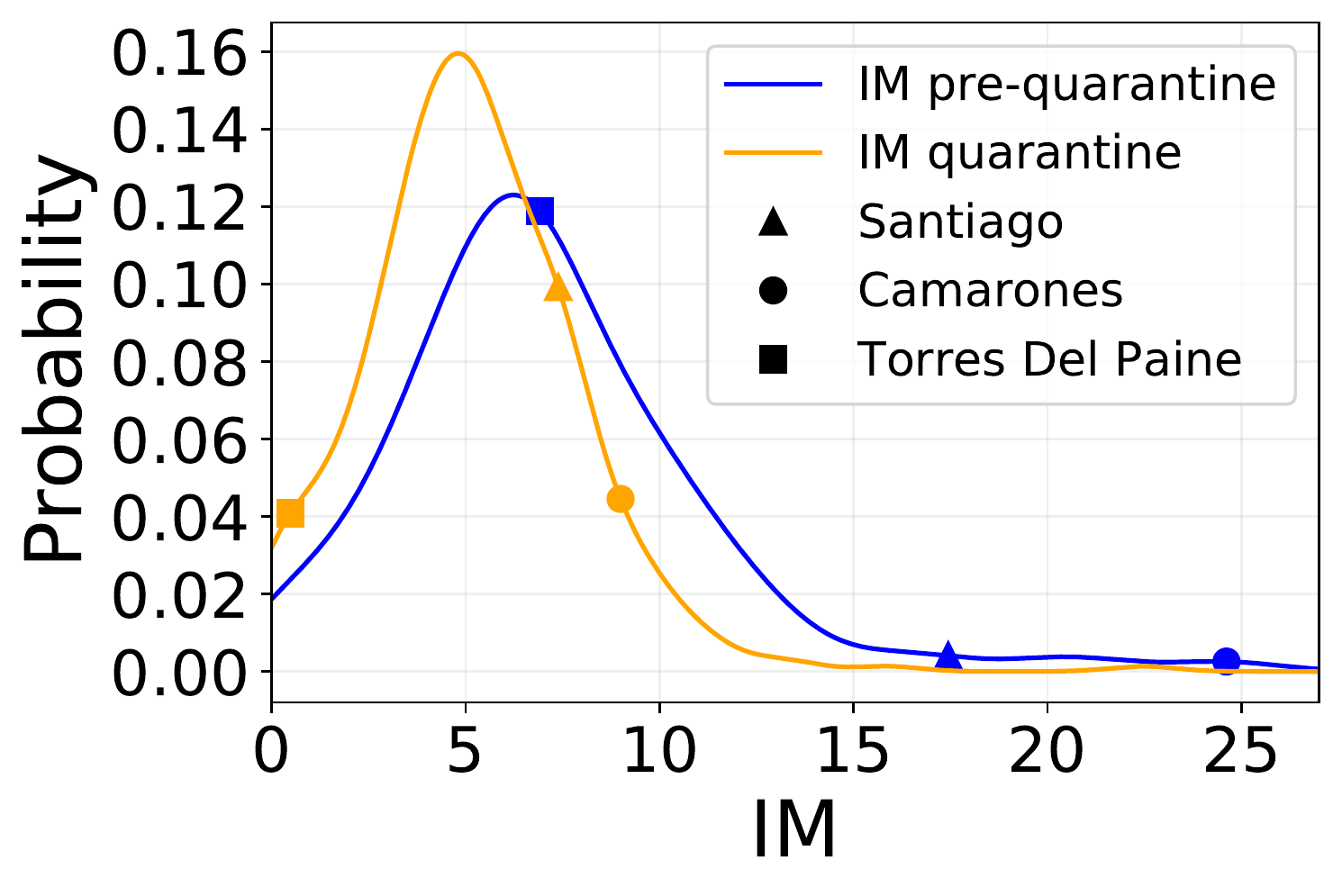}}
\subfigure[]{\label{fig:avg_IM_ext}\includegraphics[width=0.45\textwidth]{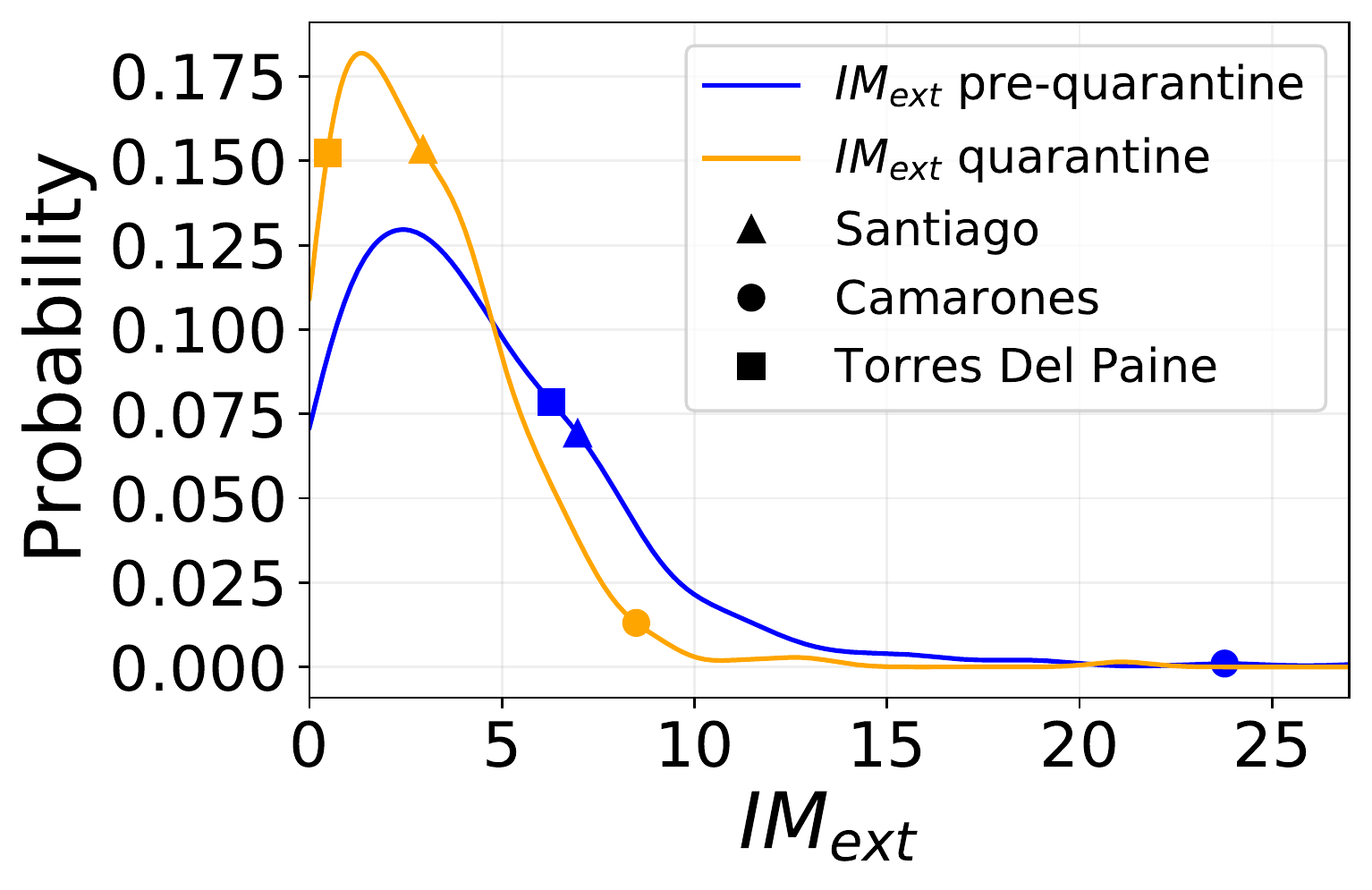}}
\subfigure[]{\label{fig:avg_IM_int}\includegraphics[width=0.45\textwidth]{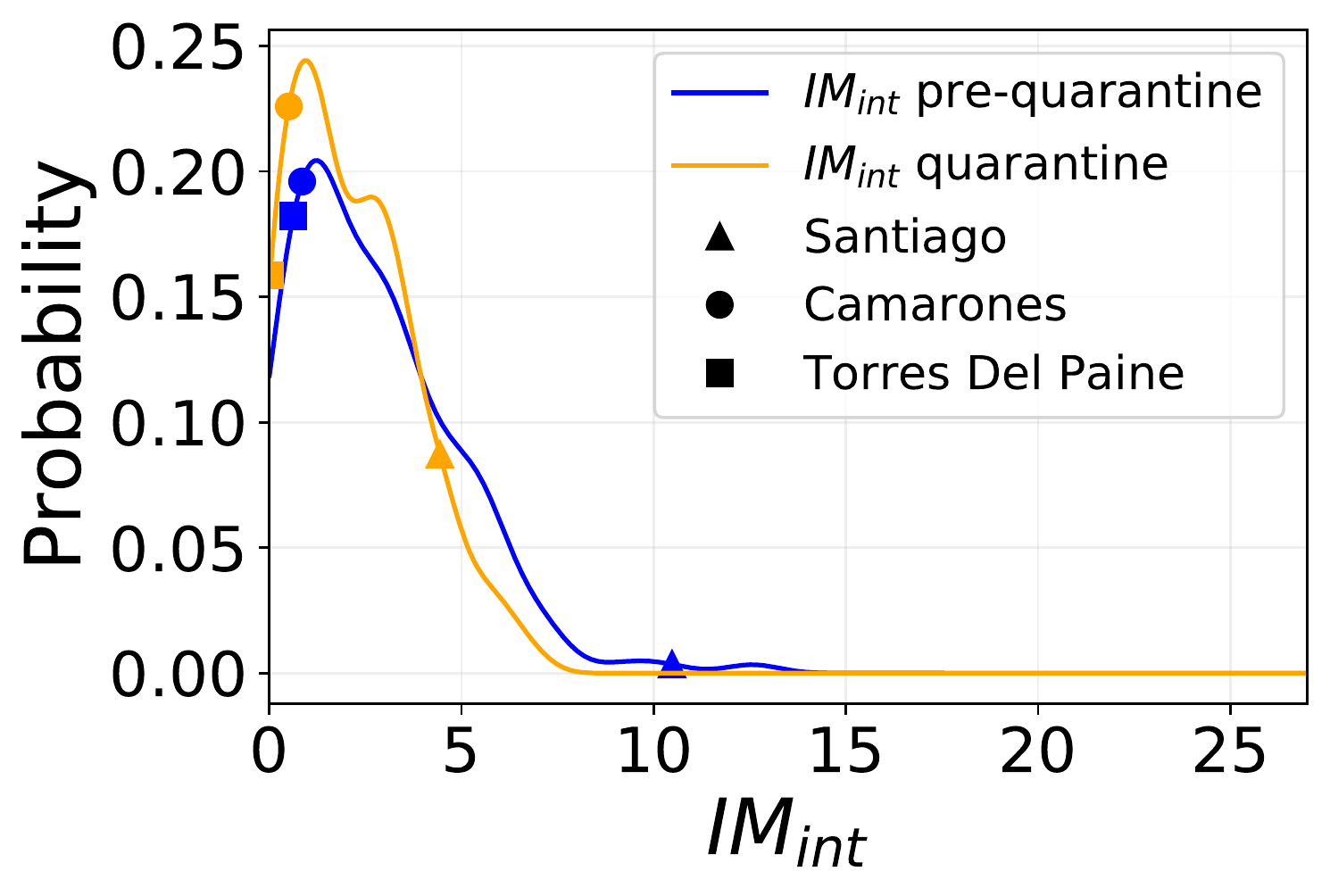}}
\subfigure[]{\label{fig:avg_IM_res}\includegraphics[width=0.45\textwidth]{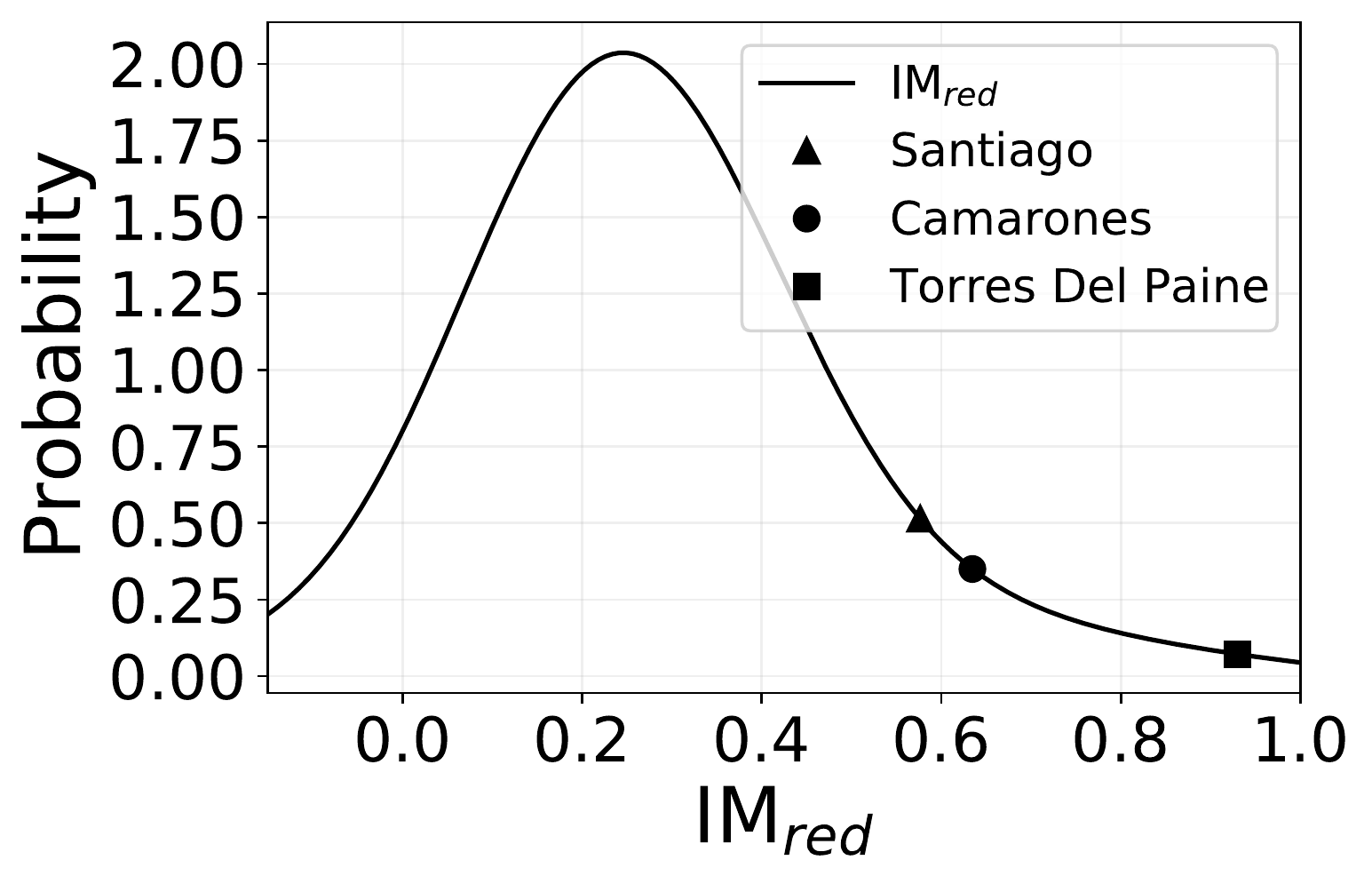}}
\caption{Distributions of IM (a),  $\mbox{IM}_{ext}$ (b) and $\mbox{IM}_{int}$ (c) for the pre-quarantine (blue) and quarantine (orange) periods, with the average values of three comunas: Santiago, Camarones and Torres Del Paine. 
(d) Distribution of $\mbox{IM}_{red}$ for all the Chileans comunas.}
\label{fig:distributions_kde}
\end{figure}

These results are supported by the distributions of the mobility indices of the two periods (Figure \ref{fig:distributions_kde}).
There is a clear shift towards the left of the distribution of the IM index (Figure \ref{fig:distributions_kde}a): \emph{(i)} the average IM during the quarantine period ($5.16 \pm 2.74$) is 27.6\% lower than the average IM during the pre-quarantine period ($7.13 \pm 4.15$); \emph{(ii)} the distribution of IM during the quarantine period is more skewed to the left, showing a decrease of the mobility in Chile during the selected days.
Regarding IM$_{int}$ and IM$_{ext}$, we observe no net shift of the curve, but rather a flattening, suggesting that intra- and inter-comuna trips decreased during the quarantine (Figures \ref{fig:distributions_kde}b and \ref{fig:distributions_kde}c).

\begin{table}[ht]\centering
\begin{tabular}{@{}cccccc@{}}
\multicolumn{6}{c}{\large Quarantine period}\\
\toprule
   & Comuna       & Region                       & IM    & IM$_ext$ & IM$_int$ \\ \midrule
1  & Rinconada    & Valparaíso                   & 22.44 & 21.09    & 1.35     \\
2  & Zapallar     & Valparaíso                   & 15.84 & 13.16    & 2.68     \\
3  & Panquehue    & Valparaíso                   & 13.30 & 11.13    & 2.17     \\
4  & Coinco       & Libertador Gen. B. O'Higgins & 13.20 & 12.36    & 0.84     \\
5  & Andacollo    & Coquimbo                     & 11.85 & 6.25     & 5.60     \\
6  & Vitacura     & Metropolitana de Santiago    & 11.33 & 4.29     & 7.04     \\
7  & Limache      & Valparaíso                   & 11.25 & 5.41     & 5.84     \\
8  & La Reina     & Metropolitana de Santiago    & 10.78 & 6.16     & 4.62     \\
9  & Concón       & Valparaíso                   & 10.75 & 4.69     & 6.06     \\
10 & Villa Alegre & Maule                        & 10.67 & 8.67     & 1.99     \\ \bottomrule
\end{tabular}
\caption{The ten comunas with the highest average value of the IM index computed over the period from June 22nd and June 28th, 2020.}
\label{tab:rank2}
\end{table}

\begin{figure}
\centering
    \subfigure[]{\label{fig:pct_quarantine}
\includegraphics[width=1\textwidth]{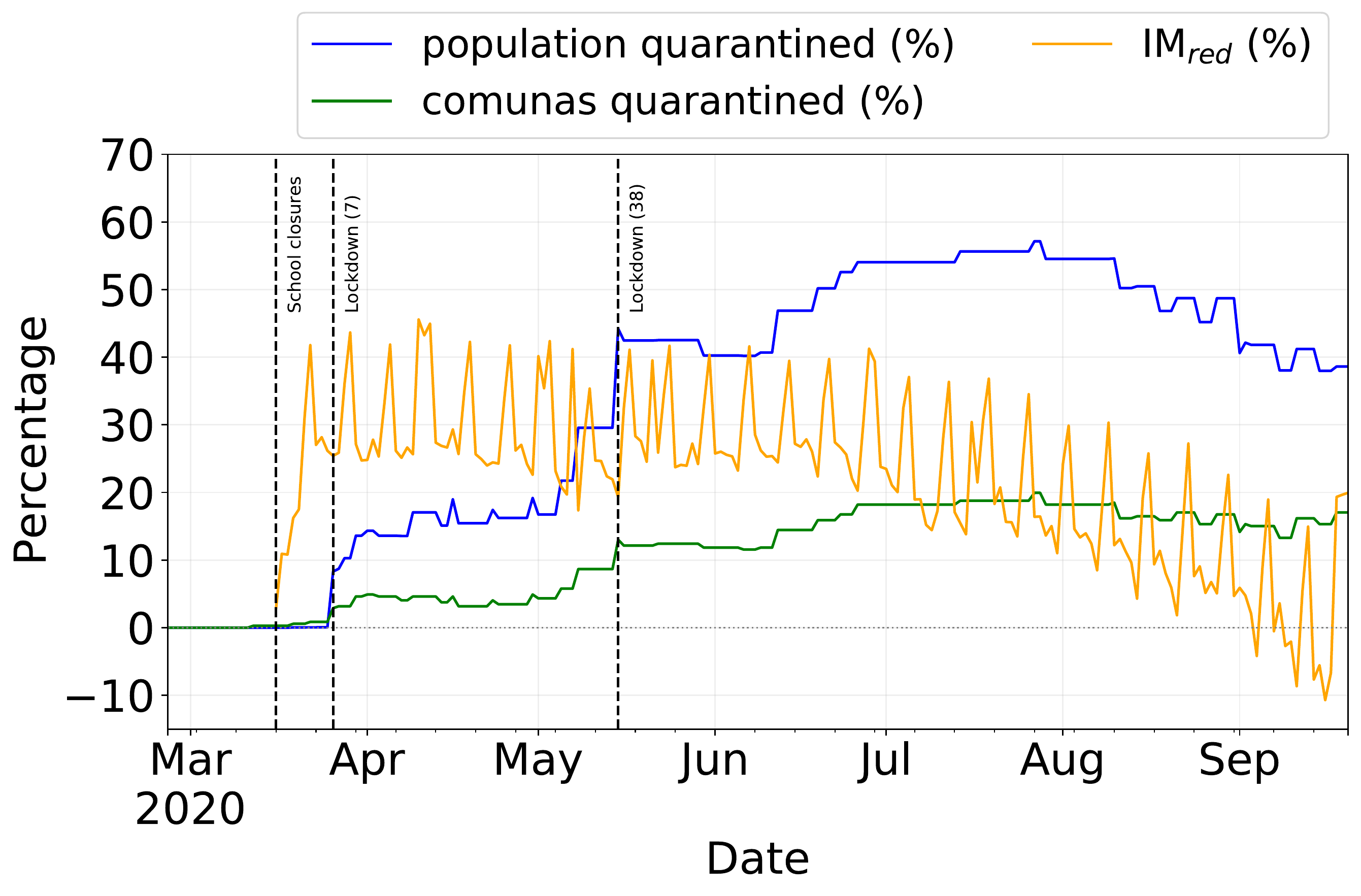}}
\hspace{2mm}
    \subfigure[]{\label{fig:quarantine_siago}
\includegraphics[width=1\textwidth]{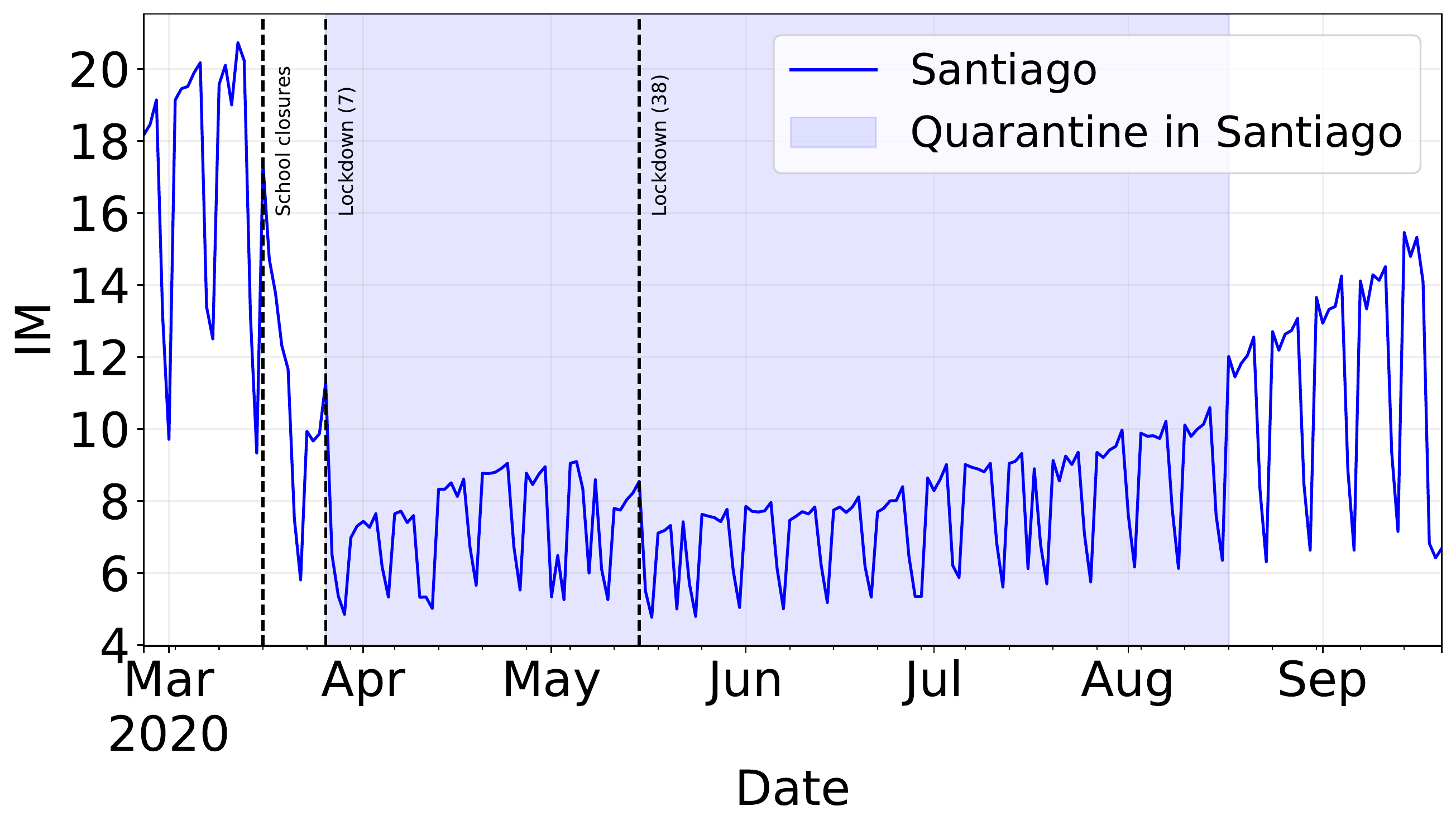}}
\caption{(a) Percentage of population under quarantine and the percentage of mobility reduction IM$_{red}$ from February 26th to September 20th, 2020.
(b) Evlution of IM index in Santiago; the blue area denotes the quarantine period.
The vertical lines denote important dates regarding NPIs in Chile;
the number in parentheses indicates the number of comunas subject to that restriction.}
\label{fig:quarantines}
\end{figure}

We further analyze the reduction of the mobility defining $\mbox{IM}_{red}$ as the relative reduction of the $\mbox{IM}$ index in the quarantine period with respect to the pre-quarantine period.
The distribution of $\mbox{IM}_{red}$ shows that a large number of comunas have a reduced mobility, following Chilean government interventions, by an average of $25.37\% \pm43.2$ (Figure \ref{fig:distributions_kde}d).
However, comunas that were not in quarantine during the quarantine period do not reduce their mobility significantly (Figure \ref{fig:quarantines}a).

The percentage of population that live in comunas where the authorities applied NPIs increases with time (Figure \ref{fig:quarantines}a) reaches its peak ($\approx$ 57\%) in late July 2020.
With the increase of the number of people under quarantine, $\mbox{IM}_{red}$ initially increases, but it slightly decreases over time even if both the number of individuals and the number of comunas under quarantine increase.
This phenomenon suggests that mobility restrictions are more effective in the short-medium term and become less effective as time goes by, and it can be observed both at regional (Figure \ref{fig:timeseriesIM}) and comuna level (Figures \ref{fig:quarantines}a and \ref{fig:quarantines}b).

\section*{Code Availability}
The code used for analysis are available at \cite{datacitation}. 
The data is also available from the general repository of the Ministry of Science of Chile at \url{https://github.com/MinCiencia/Datos-COVID19/tree/master/output/producto33}.

\section*{Acknowledgements}
Luca Pappalardo has been partially funded by EU project SoBigData++ RI, grant \#871042. Leo Ferres and Loreto Bravo thank the funding and support of Telef\'onica R\&D Chile and CISCO Chile. This research was supported by FONDECYT Grant N°1130902 to LB.

\section*{Author contributions}  
LF and LB developed and computed the mobility indices.
LP and GC made the plots and wrote the paper.
 
\section*{Competing interests}
The authors declare no competing interests.

\bibliographystyle{plain}
\bibliography{bib.bib}

\end{document}